\documentclass[letterpaper]{JHEP}
\usepackage{epsfig}

\usepackage{amssymb}

\def\drawbox#1#2{\hrule height#2pt
        \hbox{\vrule width#2pt height#1pt \kern#1pt
              \vrule width#2pt}
              \hrule height#2pt}

\def\Fund#1#2{\vcenter{\vbox{\drawbox{#1}{#2}}}}
\def\Asym#1#2{\vcenter{\vbox{\drawbox{#1}{#2}
              \kern-#2pt       
              \drawbox{#1}{#2}}}}

\def\funda{\Fund{6.5}{0.4}}
\def\asymm{\Asym{6.5}{0.4}}

\def\symm{\funda\kern-0.4pt\funda}

\usepackage{amsfonts}
\usepackage{epsfig}
\usepackage{latexsym}
\usepackage{graphicx}
\def\bequ{\begin{equation}}
\def\eequ{\end{equation}}
\def\barr{\begin{array}}
\def\earr{\end{array}}
\def\half{{1\over 2}}
\def\ben{\begin{equation}}
\def\een{\end{equation}}
\def\bena{\begin{eqnarray}}
\def\eena{\end{eqnarray}}

\renewcommand{\theequation}{\arabic{section}.\arabic{equation}}

\def\b1{e0}

\newcommand{\be}{\begin{equation}}
\newcommand{\ee}{\end{equation}}
\def\bea{\begin{eqnarray}}
\def\eea{\end{eqnarray}}
\def\Tr{\mbox{Tr}}

\def\NS{\mbox{NS}}
\def\R{\mbox{R}}

\def\ket#1{| #1 \rangle}
\def\bra#1{\langle #1 |}

\def\nn{\nonumber}

\def\half {{1 \over 2}}
\def\sgn {{sgn}}

\def\be{\begin{equation}}
\def\ee{\end{equation}}
\def\bea{\begin{eqnarray}}
\def\eea{\end{eqnarray}}

\def\lesssim{\mathrel{\hbox{\rlap{\hbox{\lower4pt\hbox{$\sim$}}}\hbox{$<$}}}}
\def\gtrsim{\mathrel{\hbox{\rlap{\hbox{\lower4pt\hbox{$\sim$}}}\hbox{$>$}}}}

\preprint{{\tt hep-th/0311068}}

\title{The Cap in the Hat: Unoriented 2D Strings and
  Matrix(-Vector) Models}   

\author{Oren Bergman and Shinji Hirano\\
Department of Physics\\
Technion, Israel Institute of Technology\\
Haifa 32000, Israel\\
\email{bergman, hirano@physics.technion.ac.il}}

\abstract{We classify the possible bosonic and Type 0 unoriented open and
closed string 
theories in two dimensions, and find their dual matrix(-vector) models.
There are no $RP^2$ R-R tadpoles in any of the models, but many of them
possess a massless tachyon tadpole.
Thus all the models we find are consistent two-dimensional string vacua,
but some get quantum corrections to their classical tachyon background.
Where possible, we solve the tadpole cancellation condition, and find
all the tachyon tadpole-free theories.}

\begin{document}

\baselineskip16pt
\parskip=4pt

\section{Introduction}

The solution of the $c=1$ matrix model in the double scaling limit 
provided an exact realization of two-dimensional string theory \cite{Gross:1990ay}.
This has been tested to high precision in perturbation theory (for a review see 
\cite{Ginsparg:is,Klebanov:ir}). 
The matrix model also predicted a non-perturbative effect 
scaling as $\exp(-1/g_s)$, which was conjectured to be a generic
feature of string theory \cite{Shenker:1990uf}. 
This conjecture was confirmed by the discovery of D-branes
in string theory \cite{Dai:ua,Polchinski:fq, Polchinski:1995mt}.
However, lacking a clear understanding of D-branes in two-dimensional
string theory, it was difficult to 
test the matrix model beyond perturbation theory
(for some recent attempts see \cite{Fukuma:1996hj}).
Furthermore, the matrix model created some problems of its own, such as the
tunneling instability \cite{Polchinski:1994jp}. 

A breakthrough was made by Zamolodchikov and Zamolodchikov,
who recently constructed the D0-brane state in Liouville theory
\cite{Zamolodchikov:2001ah}.
In light of this, 
McGreevy and Verlinde proposed to 
identify the $c=1$ matrix model with the quantum mechanics 
of the open string tachyon on $N$ unstable D0-branes of the corresponding
two-dimensional string theory \cite{McGreevy:2003kb}.
This conjecture was made precise by the
authors of \cite{Klebanov:2003km}. 
The equivalence of the $c=1$
matrix model and two-dimensional string theory is now understood
as a holographic open/closed string duality, similar to AdS/CFT.
It is also related to another recent development in string theory,
namely the condensation of open string tachyons 
(for a review see \cite{Sen:1999mg}), and the rolling tachyon
background \cite{Sen:2002nu}. 
It is remarkable to see two recent eminent ideas blend
into a paradigm, albeit in the realm of a simplified model of string theory. 

In a further development, a new interpretation of the $c=1$ matrix model
was proposed in \cite{Takayanagi:2003sm,Douglas:2003up}. 
When the potential of the matrix model is filled symmetrically,
it is conjectured to be dual to the two-dimensional Type 0B string theory.
This avoids the tunneling
instability of \cite{Polchinski:1994jp}, leading to an unambiguously
defined non-perturbative string theory. Many other recent developments 
have appeared in \cite{Martinec:2003ka}-\cite{Alexandrov:2003un}.

In this paper, we will extend these conjectures to unoriented
open and closed string theory.
We will classify the possible unoriented bosonic and Type 0 string theories 
in two dimensions (the former was already studied in \cite{Nakayama:2003ep}),
solve the tadpole cancellation condition for each one,
and construct the corresponding matrix(-vector) models using D0-branes.
Our results for the tadpole-free theories are summarized in table 1. 
More general theories can be found in table 2,
and the corresponding matrix models in table 3.

The paper is organized as follows.
In section 2 we review the unoriented bosonic string.
In section 3 we classify all possible two-dimensional unoriented
open and closed Type 0 strings.
In section 4 we derive the corresponding matrix models by
analyzing the D0-branes in each theory, and 
section 5 contains our conclusions.

\medskip

As this paper was being written, the paper \cite{Gomis:2003vi} appeared, 
with which there is some overlap.

\begin{table}
\begin{center}
\begin{tabular}[h]{|c|c|c|c|}
\hline
theory & closed & D1-brane & D0-brane matrix(-vector) model\\
       & strings & open strings & \\
\hline
 &&& \\[-10pt]
bos/$\Omega_-$ & $T$ & 
$Sp(1) ,\,  {\bf 1}$ & $Sp({n_0\over 2}) ,\,  \asymm + 2\,\funda$ \\[10pt]
\hline
&&&\\[-10pt]
0B/$\Omega_\pm$ & $T$ & -- &
$
\begin{array}{l}
SO(n_0),\, \asymm_{\,b}\\[5pt]
Sp({n_0\over 2}) ,\, \symm_{\,b}
\end{array}
$
\\[20pt]
\hline
&&&\\[-10pt]
0B/$\widehat{\Omega}_-$ & $T, C_0$ & 
$
\begin{array}{l}
 U(2) ,\, {\bf 1}_b^0 \\[5pt]
 U(1)^2 ,\, 2(\pm,\mp)_f 
\end{array}
$
&
$
Sp({n_0\over 2}) ,\,  \asymm_{\,b} +
\left\{
\begin{array}{l}
4\,\funda_{\,b\,or\, f} \\[5pt]
2\,\funda_{\,b} + 2\,\funda_{\,f}
\end{array}\right.
$\\[20pt]
\hline
&&&\\[-10pt]
0A/$\Omega_-$ & $T, C_1^+$ &
$
\begin{array}{l}
SO(2) ,\, {\bf 1}_b\\[5pt]
- ,\, {\bf 1}_f
\end{array}
$ &
$U(n_0) ,\, \asymm_{\,b}  + 
\left\{
\begin{array}{l}
2[\funda 
+ \,\overline{\funda}]_{\,b\, or\, f}\\[5pt]
[\funda + \,\overline{\funda}]_{\,b} 
+ [\funda + \,\overline{\funda}]_{\,f}
\end{array}\right.$
\\[10pt]
\hline
&&&\\[-10pt]
0A/$\widehat{\Omega}_-$ & $T, C_1^-$ &
$Sp(1) ,\, {\bf 1}_b$ & 

$ Sp({n_0\over 2})\times Sp({\bar{n}_0\over 2}) , (\funda,\funda)_{\,b} 
+ 2[(\funda,{\bf 1}) 
+ ({\bf 1},\funda)]_{b\, or \, f}$
\\[20pt]
\hline
\end{tabular}
\end{center}
\caption{Tadpole-free unoriented open and closed string theories in two 
dimensions and their dual matrix(-vector) models.}
\label{tadpole_free_models}
\end{table}

\section{Unoriented bosonic string}

The two-dimensional unoriented bosonic string was analyzed recently in
\cite{Nakayama:2003ep}. Let us review this model and clarify some points.
As is well known, gauging the world-sheet parity symmetry 
of the 26 dimensional critical bosonic string theory gives
an unoriented string theory, which has a massless tadpole
(for the dilaton and graviton) on $RP^2$. The easiest way 
to compute this tadpole is from the one loop Klein bottle amplitude.
Channel duality relates this to a tree amplitude involving the
square of the $RP^2$ tadpole. 
This tadpole can be cancelled against a disk tadpole by introducing
open strings.
By comparing the Klein bottle,
M\"obius strip and cylinder amplitudes one finds that cancellation
of the massless tadpole requires an open string Chan-Paton gauge group $SO(2^{13})$.
However one is still faced with the problem posed by the closed string tachyon.


In two dimensions the situation is somewhat better, in that the ground state
is massless (though we continue to call it a tachyon), and in fact
is the only propagating state.
It is the tadpole of this state which will concern us.
In non-critical string theory one of the dimensions is a 
Liouville field, which complicates the 
computation of the one loop amplitudes due to the
presence of the bulk and boundary Liouville interactions 
(cosmological constants).
However, the massless tadpoles can be obtained from the one-loop
amplitudes in the {\em free} field theory, which are easily computable.
This is because the massless tadpole in Liouville theory corresponds to the IR pole
of the corresponding one-point function (on the disk or $RP^2$), and the residue
is given by the one-point function in the free theory 
\cite{Fateev:2000ik, Zamolodchikov:1995aa}.
A heuristic way to illustrate this is to look at just the Liouville zero mode, and treat
the bulk interaction perturbatively. In this approximation the one-point     
function on a Riemann surface with Euler number $\chi$ is given by
\begin{eqnarray}
\langle e^{\alpha\phi}\rangle_\chi
&=&\int_{-\infty}^{\infty}d\phi_0\, e^{\alpha\phi_0}
e^{-\chi Q\phi_0-\mu e^{2b\phi_0}-\mu_Be^{b\phi_0}}\nn\\
&=&{1\over b}\mu^{-{1\over 2b}(\alpha-\chi Q)}
\sum_{k=0}^{\infty}{(-1)^k\over k!}
\left({\mu\over\mu_B^2}
\right)^{{\alpha-\chi Q\over 2b} +k}
\Gamma\left({\alpha-\chi Q\over b}+2k\right)\;,
\label{zmpole}
\end{eqnarray}
where $\alpha = Q + 2iP$.
%
The disk and crosscap ($RP^2$) both have $\chi=1$,
so the massless tadpole corresponds to the $P\rightarrow 0$ pole of the 
Gamma function at the $k=0$ order. The residue of this pole is independent
of $\mu$ and $\mu_B$, and corresponds to the free field result.
%
%

The Klein bottle, cylinder and M\"obius amplitudes (per unit volume) 
found in \cite{Nakayama:2003ep} 
are (in the tree channel)
\begin{eqnarray}
 {\cal A}_K &=& \int_0^\infty {ds\over 8\pi^3} \\
 {\cal A}_C &=& n^2 \int_0^\infty {ds\over 32\pi^3} \\
 {\cal A}_M &=& \pm n \int_0^\infty {ds\over 8\pi^3} \;,
\end{eqnarray}
where the sign in the M\"obius amplitude refers to the choice we have on 
how $\Omega$ acts on the CP factor. We will denote this choice by $\Omega_\pm$,
in correspondence with the sign of the $RP^2$ tachyon tadpole. 
The corresponding CP groups are $SO(n_1)$ in the $\Omega_+$ case, and
$Sp(n_1/2)$ in the $\Omega_-$ case.\footnote{Note that this is the opposite
convention to the one usually used in critical string theory.
There the sign of $\Omega_\pm$ refers to the sign of the $RP^2$ dilaton tadpole
(and also the RR tadpole for the superstring), which is the opposite
of the $RP^2$ tachyon tadpole.}
Since the open string ground state is even under world-sheet parity $\Omega$,
the massless open string tachyon is a symmetric matrix in the
orthogonal case and a skew-symmetric matrix in the symplectic case.
The latter corresponds to the antisymmetric representation of the symplectic group.
In the tadpole-free case we therefore get $Sp(1)$ with a tachyon in the ${\bf 1}$.

This is also consistent with what is obtained in the exact Liouville calculation
\cite{Nakayama:2003ep}, using the explicit form of the Liouville crosscap state
derived in \cite{Hikida:2002bt}.

\begin{table}
\begin{center}
\begin{tabular}[h]{|c|c|l|l|c|}
\hline
theory & closed strings & open string CP group& open string matter 
  & $T$ tadpole \\
\hline
 &&&& \\[-10pt]
bos/$\Omega_\pm$ & $T$ & $
\begin{array}{l}
SO(n_1) \\[5pt]
Sp({n_1\over 2})
\end{array}$ 
& 
$
\begin{array}{l}
 \symm \\[5pt]
 \asymm
\end{array}$ 
&
$
\begin{array}{c}
n_1+2 \\[5pt] 
n_1-2
\end{array}$ \\[20pt]
\hline
&&&&\\[-10pt]
0B/$\Omega_\pm$ & $T$ & 
$
\begin{array}{l}
SO(n_1^+)^2\times SO(n_1^-)^2 \\[5pt]
Sp({n_1^+\over 2})^2\times Sp({n^-_1\over 2})^2
\end{array}
$ 
& 
$
\begin{array}{l}
(n_1^\pm,n_1^\pm)_b + (n_1^\pm,n_1^\mp)_f\\[5pt]
(n_1^\pm,n_1^\pm)_b + (n_1^\pm,n_1^\mp)_f
\end{array}
$ 
& 
$
\begin{array}{c}
2n_1 \\[5pt]
2n_1
\end{array} $ \\[20pt]
\hline
&&&&\\[-10pt]
0B/$\widehat{\Omega}_\pm$ & $T, C_0$ & 
$
\begin{array}{lcl}
 U(n^+_1)\times U(n^-_1) \\[5pt]
 U(n^+_1)\times U(n^-_1)
\end{array}
$
& 
$
\begin{array}{l}
 \symm^\pm_{\,b} + 2(n^\pm_1,\bar{n}^\mp_1)_f \\[5pt]
 \asymm^\pm_{\,b} + 2(n^\pm_1,\bar{n}^\mp_1)_f
\end{array}
$
& 
$
\begin{array}{c}
n_1 + 2 \\[5pt]
n_1 - 2
\end{array} $ \\[20pt]
\hline
&&&&\\[-10pt]
0A/$\Omega_\pm$ & $T, C_1^+$ &
$
\begin{array}{l}
 Sp({n_1^+\over 2}) \times Sp({n_1^-\over 2}) \\[5pt]
 SO(n_1^+)\times SO(n_1^-) 
  \end{array}
$
& 
$
\begin{array}{l}
 \symm^\pm_{\,b} + 2(n_1^+,n_1^-)_f \\[5pt]
 \asymm^\pm_{\,b} + 2(n_1^+,n_1^-)_f
\end{array}
$
& 
$
\begin{array}{c}
n_1+2 \\[5pt] 
n_1-2
\end{array}$ \\[20pt]
\hline
&&&&\\[-10pt]
0A/$\widehat{\Omega}_\pm$ & $T, C_1^-$ &
$
\begin{array}{l}
 SO(n_1^+)\times SO(n_1^-) \\[5pt]
 Sp({n_1^+\over 2}) \times Sp({n_1^-\over 2}) 
\end{array}
$
& 
$
\begin{array}{l}
 \symm^\pm_{\,b} + 2(n_1^+,n_1^-)_f \\[5pt]
 \asymm^\pm_{\,b} + 2(n_1^+,n_1^-)_f
\end{array}
$
& 
$
\begin{array}{c}
n_1+2 \\[5pt] 
n_1-2
\end{array}$ \\[20pt]
\hline
\end{tabular}
\end{center}
\caption{Unoriented open and closed string theories in two dimensions.}
\label{unoriented_strings}
\end{table}

\section{Unoriented Type O strings}

\subsection{Review of critical Type 0 strings}

Let us first review the critical Type 0 string theories and
their orientifolds. Many of the properties we will review
are shared by the two-dimensional Type 0 theories.
The critical Type 0 string theories are 
ten-dimensional modular invariant
closed fermionic string theories, which consist of an NS-NS 
and R-R sector only, with a diagonal GSO projection:
$$
\begin{array}{rl}
 \mbox{Type 0A:} & \;\;(\NS+,\NS+) + (\NS-,\NS-) + (\R+,\R-) + (\R-,\R+)\\[5pt]
 \mbox{Type 0B:} & \;\;(\NS+,\NS+) + (\NS-,\NS-) + (\R+,\R+)+(\R-,\R-)\;.
\end{array}
$$
As such, they are not spacetime supersymmetric, and in fact possess a 
tachyon in the $(\NS-,\NS-)$ sector.
The Type 0 theories also contain two sets of massless R-R gauge fields,
$C_p^+$ and $C_p^-$ (these are even and odd linear combinations of the 
fields in the above sectors),
and correspondingly two sets of R-R charged D-branes, $D(p+1)_{\eta=\pm}$,
with $p$ even in Type 0B and odd in Type 0A 
\cite{Bergman:1997rf,Klebanov:1998yy,Bergman:1999km}.
As in the Type II theories, the Type 0 theories also possess uncharged
and unstable D-branes of the ``wrong'' dimensions. They also come in
two varieties, and are denoted $\widetilde{Dp}_\eta$.
The corresponding D-brane boundary states are given by
\begin{eqnarray}
\ket{D(p+1)_\eta} &=& {1\over\sqrt{2}}\left(
  \ket{B(p+1),\eta}_{NSNS} + \ket{B(p+1),\eta}_{RR}\right) \nonumber\\
\ket{\widetilde{Dp}_\eta} &=& \ket{Bp,\eta}_{NSNS}\;.
\label{D-branes}
\end{eqnarray}
It follows immediately from channel duality (Appendix A) that open strings
between D-branes of the same sign $\eta$ are spacetime bosons, and those between 
D-branes of opposite sign are spacetime fermions.
Furthermore, the open string spectrum of the charged D-branes
contains only GSO-even states,
whereas that of the neutral D-branes contains both GSO-even and GSO-odd
states (and therefore a tachyon in the same sign case).

A number of different unoriented string theories can be obtained
from the Type 0 theories \cite{Bianchi:1990yu,Bergman:1999km}. 
These correspond to gauging the
discrete symmetries $\Omega$, 
$\widehat{\Omega}=\Omega\cdot(-1)^{F_L}$, where $F_L$ is the left-moving
part of the spacetime fermion number (or NSR parity), and 
$\Omega'=\Omega\cdot(-1)^f$, where $f$ is the left-moving world-sheet
fermion number.
In Type 0B all three symmetries are involutions,
and in Type 0A the first two are involutions, whereas $\Omega'$
generates a $Z_4$ symmetry.
This gives five distinct models: 0B$/\Omega$, 0B$/\widehat{\Omega}$,
0B$/\Omega'$, 0A$/\Omega$ and 0A$/\widehat{\Omega}$.
The sixth possibility, 0A$/\Omega'$, is actually equivalent to one of the
Type 0B models.

The models differ in their perturbative and D-brane spectra,
as well as in the $RP^2$ tadpoles they possess.
By channel duality for the Klein bottle (table A.3)
we can immediately see that because of the diagonal GSO projection 
$(1+(-1)^{f+\tilde{f}})$ in the loop channel, 
the $\Omega$ models have an NS-NS tadpole,
but no R-R tadpole.
The same is true for the $\widehat{\Omega}$ models,
since the effect of $(-1)^{F_L}$ is just to change the sign
of the R-R contribution in the trace.
On the other hand, the effect of $(-1)^f$ is to change the diagonal
GSO projection to $((-1)^f+(-1)^{\tilde{f}})$, therefore the 
$\Omega'$ model possesses an R-R tadpole, but no NS-NS 
tadpole.\footnote{Interestingly, this theory is tachyon-free, since the tachyon
is odd under $(-1)^f$, and therefore under $\Omega'$.}
Only this model {\em requires} the addition of open strings for 
consistency, resulting in the Chan-Paton gauge group $U(32)$.
However this introduces NS-NS disk tadpoles, including a tachyon tadpole,
as well as a massless (dilaton) tadpole.

We can determine the states of the $RP^2$ NS-NS tadpoles in the $\Omega$
and $\widehat{\Omega}$ models using channel duality. From the action of 
$\Omega$ on the NS-NS and R-R ground states of Type 0B
\cite{Gimon:1996rq},
\begin{eqnarray}
 \Omega\ket{0}_{NS}\otimes\ket{0}_{NS} &=& \ket{0}_{NS}\ket{0}_{NS}\nonumber \\
 \Omega\ket{S^\alpha}_R\otimes\ket{\widetilde{S}^\beta}_R &=& 
        - \ket{S^\beta}_R\otimes\ket{\widetilde{S}^\alpha}_R  \;,
\end{eqnarray}  
we see that the Klein bottle amplitude has the form
\be
 \bra{C,+}\Delta\ket{C,-}_{NSNS} \mp \bra{C,+}\Delta\ket{C,+}_{NSNS}
\ee
in the 0B$/\Omega$ and 0B$/\widehat{\Omega}$ theories, respectively.
The tachyon contribution cancels in the first case and adds in the second
case. Therefore the 0B$/\Omega$ model has a dilaton tadpole but no
tachyon tadpole, whereas the 0B$/\widehat{\Omega}$ model has a tachyon
tadpole but no dilaton tadpole.
In Type 0A the action of $\Omega$ on the R-R ground states is
\be
 \Omega\ket{S^\alpha}_R\otimes\ket{\widetilde{S}^{\dot{\beta}}}_R = 
        - \ket{S^{\dot{\beta}}}_R\otimes\ket{\widetilde{S}^\alpha}_R \;,
\ee
or in other words
\be
 \Omega: C^{\pm}_p \rightarrow \pm C^{\pm}_p\;.
\ee
The contributions of $C^+_p$ and $C^-_p$ therefore cancel in the R-R
trace, and the Klein bottle in Type 0A is just
\be
 \bra{C,+}\Delta\ket{C,-}_{NSNS} \;.
\ee
The 0A $\Omega$ and $\widehat{\Omega}$ models therefore have both
a tachyon and a dilaton tadpole.

The main problem with these models, as with the oriented Type 0 strings,
is the presence of the tachyon (except in the $\Omega'$ model). 
Ignoring this for the moment (or assuming the tachyon has rolled 
down to its true vacuum), we are faced with massless NS-NS tadpoles
in 0B$/\Omega$, 0A$/\Omega$ and 0A$/\widehat{\Omega}$.
One has two options here: cancel the NS-NS tadpole by adding open strings
(in such a way as not to introduce a R-R tadpole), or employ the 
Fischler-Susskind mechanism. The first option yields the CP gauge groups 
$(SO(n)\times SO(32-n))^2$ and $SO(n)\times SO(32-n)$ for the 0B and (both) 0A
models, respectively.

\subsection{Two-dimensional Type 0 strings}

Two-dimensional Type 0 strings were described in 
\cite{Takayanagi:2003sm,Douglas:2003up}.
The relevant two-dimensional CFT combines a free $N=1$ SCFT corresponding
to time, and an $N=1$ super-Liouville theory corresponding to space.
The latter is given by
\be
 S_L  = {1\over 4\pi}\int d^2z
       \left(\partial\phi\bar{\partial}\phi + \psi\bar{\partial}\psi
      + \bar{\psi}\partial\bar{\psi} 
      + 2i\mu_0b^2\psi\bar{\psi}e^{b\phi}
      + \mu_0^2b^2e^{2b\phi}\right) \;,
\ee
and there is an implicit background charge given by $Q\chi$, where $\chi$ is
the Euler number of the world-sheet, and $Q$ is given by
\be
 Q = b +{1\over b} \;.
\ee
To get two dimensions we take the limit $b\rightarrow 1$, keeping
$\mu = \mu_0\gamma((1+b^2)/2)/4$ fixed, where $\gamma(x)=\Gamma(x)/\Gamma(1-x)$.
The resulting two-dimensional string theories have no propagating physical 
excitations\footnote{There are however discrete physical states
\cite{Bouwknegt:1991yg} (see also \cite{Itoh:1991ix}).}, and 
the only propagating fields are the ground states. In the NS-NS sector
this is the tachyon, which is actually massless due to the background charge. 
The R-R sector contains a single massless scalar $C_0$ (combining the self-dual
and anti-self-dual parts) in the Type 0B case,
and a pair of massless vectors $C_1^\pm$ in the Type 0A case.\footnote{There
is an imaginary shift $ib$ of the background charge in the R-R sector due
to the presence of the fermionic zero modes.} 
The considerations of world-sheet fermion
spin structures are basically the same as in the ten-dimensional case,
so many of the results we described above hold also in the two-dimensional
case. The main difference (in the closed string case) is that the Liouville
interaction breaks the global $(-1)^f$ symmetry.\footnote{This is why it cannot
be gauged, for $\mu_0\neq 0$, to yield two-dimensional Type II strings.}
Consequently the $\Omega' = \Omega\cdot (-1)^f$ model does not exist in
two-dimensions.\footnote{We thank Jaume Gomis for pointing this out to us.}
We are left with just the two $\Omega$ models, and the two
$\widehat{\Omega}$ models.

Open strings can be incorporated by including boundaries with either
Neumann or Dirichlet boundary conditions for the matter and
super-Liouville fields. 
The problem of a Neumann boundary condition for the Liouville field was
solved in \cite{Fateev:2000ik,Teschner:2000md}, and the Dirichlet case
was solved in \cite{Zamolodchikov:2001ah}.
These were later extended to the super-Liouville theory in 
\cite{Fukuda:2002bv,Ahn:2002ev}. This requires the addition of a boundary
Liouville action \cite{Douglas:2003up}
\be
 S_{\partial L} = {1\over 2\pi}\oint dx
    \left(\gamma\partial_x\gamma 
     - \mu_Bb\gamma(\psi + \eta\bar{\psi})e^{b\phi/2}
     + \mu_B^2e^{b\phi}\right)\;,
\ee
where $\gamma$ is a boundary fermionic field.

The Liouville Neumann boundary state is labeled by a parameter $\nu$,
which is defined by
\be
 \cosh^2(\pi b\nu) = {\mu_B^2\over 2\mu_0} \cos\left({\pi b^2\over 2}\right)\;,
\ee
and can have four possible spin structures labeled by $\eta$ and $\eta'$,
where $\eta'=\pm 1$ corresponds to the R-R and NS-NS sector, respectively.
In terms of the Ishibashi states $\ket{B,P,\eta}_{\eta'}$, the Cardy
boundary states are given by
\be
 \ket{B;\nu,\eta}_{\eta'} = \int_0^\infty dP\, \Psi_\eta^{\eta'}(\nu,P)
    \ket{B;P,\eta}_{\eta'} \;,
\ee
where the different boundary state wave-functions 
(normalized disk one-point functions) are given by 
\begin{eqnarray}
\Psi^{NS}_{\eta}(\nu,P)&=&(2\mu)^{-iP/b}
\frac{\Gamma(1+iPb)\Gamma(1+iP/b)} 
{-\pi iP}\cos(2\pi P\nu)\ ,\\
\Psi^{RR}_{\eta=+\sgn(\mu)}(\nu,P)&=&(2\mu)^{-iP/b}
\frac{\Gamma(\half+iPb)\Gamma(\half+iP/b)} 
{\sqrt{2}\pi}\cos(2\pi P\nu)\ ,\\
\Psi^{RR}_{\eta=-\sgn(\mu)}(\nu,P)&=&(2\mu)^{-iP/b}
\frac{\Gamma(\half+iPb)\Gamma(\half+iP/b)}    
{\sqrt{2}\pi}\sin(2\pi P\nu)\ .
\end{eqnarray}
Therefore Type 0B contains two kinds of R-R charged D1-brane $D1_\pm$,
and Type 0A contains two kinds of neutral D1-brane $\widetilde{D1}_\pm$,
as in (\ref{D-branes}). 
The open string spectrum has no physical states for two Type 0B D1-branes
of the same sign (since the tachyon is GSO-projected out), and
a single massless fermion for two D1-branes of the opposite sign.
For the Type 0A D1-branes, the physical open string spectrum contains
a massless tachyon in the same sign case, and a pair of massless fermions
in the opposite sign case (no GSO-projection).

We defer the discussion of the Dirichlet boundary states,
{\em i.e.} the D0-branes, to the next section, where we will
construct the matrix models.
Let us next analyze the four possible unoriented models in turn.
We will apply what we have learned in the ten-dimensional models
using channel duality of the world-sheet fermionic spin structures,
and also verify the results by explicit computation of the relevant 
amplitudes in the two-dimensional models. 
The results are summarized in table 2.

\subsection{0B/$\Omega$}

The R-R scalar $C_0$ is odd under $\Omega$, and the tachyon $T$ is even,
so the tachyon is the only propagating degree of
freedom in this theory. The non-dynamical R-R two-forms $C_2^\pm$ are even,
so both $D1_\eta$-branes and $\overline{D1}_\eta$-branes are invariant.
Exactly the same world-sheet spin structure considerations as in the ten-dimensional
case show that there is no $RP^2$ R-R tadpole, and that there is no tachyon
tadpole.\footnote{This was originally pointed out to us independently by Jaume Gomis.}
We can also demonstrate this by an explicit computation of the Klein
bottle amplitude in the {\em free} field theory\footnote{Our convention 
for the open string
  momentum is fixed by   
$L_0=\half\left(p_0^2+p_1^2\right)+\cdots$,
 while the closed string momentum (squared) differs by a factor of 
${1\over 4}$, thus $L_0={1\over 8}(p_0^2+p_1^2)+\cdots$ and 
$\widetilde{L}_0={1\over 8}(p_0^2+p_1^2)+\cdots$.} :
\begin{eqnarray}
 {\cal A}_K &=& \half\int_0^{\infty}{dt\over 2t}
\Tr_{NSNS+RR}\left({1+(-1)^{f+\tilde{f}}\over 2}
\Omega\,e^{-2\pi t(L_0+\widetilde{L}_0)} \right) \nn\\
  &=& \half\int_0^{\infty}{dt\over 2t}
\int{d^2p\over(2\pi)^2}e^{-{\pi t\over 2}(p_0^2+p_1^2)}
\Tr_{NSNS+RR}
\Biggl({1+(-1)^{f+\tilde{f}}\over 2}\Omega 
\Biggr) \;.
\end{eqnarray}
Once we ignore the Liouville
interaction the actual calculation becomes trivial, since all the
excited states in two-dimensional string theory are longitudinal 
(except for non-propagating discrete states) and
thus cancelled out by the ghost contributions. After the usual modular
transformation to the tree channel $s=\pi/2t$ we find
\be
 {\cal A}_K = {2\over (2\pi)^3}\int_0^{\infty}ds\, (1-1)
   = \int_0^\infty ds\, \langle T\rangle_{O1}^2\ ,
\ee
and therefore $\langle T\rangle_{O1}=0$. 
This model is therefore tadpole-free without open strings.

Nevertheless, one can consider adding open strings by including
$D1_\eta\overline{D1}_\eta$ pairs. This does not introduce a net R-R
charge, so the theory is consistent, but it will have a disk tachyon
tadpole. We can compute this tadpole from the NS-NS exchange part of 
the cylinder amplitude.
Channel duality tells us that this is 
\be
{\cal A}_C^{NSNS} = {n_1^2\over 2}\int_0^{\infty}{dt\over 2t}
\,\Tr_{NS}\,\half e^{-2\pi tL_0}
 = {n_1^2\over 4}\int_0^{\infty}{dt\over 2t}\,
  \int {d^2 p\over (2\pi)^2}e^{-\pi t(p_0^2+p_1^2)}\;,
\ee
where again the free field computation reduces to just the massless tachyon
contribution.
In the tree channel this becomes (with $s=\pi/t$)
\be
 {\cal A}_C^{NSNS} = {n_1^2\over 4(2\pi)^3} \int_0^\infty  ds 
   = \int_0^\infty  ds\,\langle T\rangle_{D1_\eta}^2 \;,
\ee
and so $\langle T\rangle_{D1_\eta}=n_1/(2(2\pi)^{3/2})$.
For completeness we also present the M\"obius strip amplitude:
\begin{eqnarray}
{\cal A}_M&=&
\half\int_0^{\infty}{dt\over 2t}
\,\Tr_{NS}\,\left(\half\left(1+(-1)^{f}\right){\Omega}\,
e^{-2\pi tL_0}\right) \nn\\
 &=& \pm n_1\,\half\int_0^{\infty}{dt\over 2t}
\,\int {d^2 p\over (2\pi)^2}e^{-\pi t(p_0^2+p_1^2)}
(-i)\half(1-1)\;.
\end{eqnarray}
The $-i$ comes from the action of $\Omega$ on the NS ground state.
In the tree channel with $s=\pi/4t$ this becomes
\be
{\cal A}_M = \pm {2n_1\over (2\pi)^3 }\int_0^{\infty}ds\,(-i)\half(1-1)
 = 2\int_0^\infty ds\, \langle T\rangle_{D1_\eta}\langle T\rangle_{O1}\;.
\label{0Bomegamobius}
\ee
The open string spectrum for $n_1^+$ $D1_+\overline{D1}_+$ pairs
and $n_1^-$ $D1_-\overline{D1}_-$ pairs contains massless tachyons
in the bi-fundamentals $(n_1^+,n_1^+) + (n_1^-,n_1^-)$ of either
$SO(n_1^+)^2\times SO(n_1^-)^2$ or $Sp(n_1^+/2)^2\times Sp(n_1^-/2)^2$,
and massless fermions in the bi-fundamentals
$2(n_1^+,n_1^-)$.


\subsection{0B/$\widehat{\Omega}$}

As in ten-dimensions, this model does not have a R-R tadpole either.
It does however have a massless tachyon tadpole, which we will compute below. 
In this case $C_0$ is even and $C_2^\pm$ is odd, 
since $(-1)^{F_L}$ gives an additional minus sign for all R-R fields.
Therefore both $T$ and $C_0$ survive in the closed string spectrum.
On the other hand, since $C_2^\pm$ is odd, $\widehat{\Omega}$ interchanges
the $D1_\eta$-brane and the $\overline{D1}_\eta$-brane. 
The invariant $D1_\eta\overline{D1}_\eta$ 
combination is neutral, so we denote it $\widetilde{D1}_\eta$, by analogy
with the ``wrong'' dimension neutral D-branes.
The open string states on a single $\widetilde{D1}_\eta$
can be represented in terms of $U(2)$ Chan-Paton factors as
\begin{eqnarray}
11+\bar{1}\bar{1} \; \mbox{strings}:\;\;
\left|p;N;\mathbb{I} \rangle\right.
&=&{1\over\sqrt{2}}\,\mathbb{I}_{ij}\left|p;N;ij\rangle\right.
\label{D1tildestatesfirst}\nn\\
1\bar{1}+\bar{1}1 \; \mbox{strings}:\;\;
\left|p;N;\sigma_1 \rangle\right.
&=&{1\over\sqrt{2}}\,(\sigma_1)_{ij}
\left|p;N;ij\rangle\right.\nn\\
1\bar{1}-\bar{1}1 \; \mbox{strings}:\;\;
\left|p;N;\sigma_2 \rangle\right.
&=&{1\over\sqrt{2}}\,(\sigma_2)_{ij}
\left|p;N;ij\rangle\right.\nn\\
11-\bar{1}\bar{1} \; \mbox{strings}:\;\;
\left|p;N;\sigma_3 \rangle\right.
&=&{1\over\sqrt{2}}\,(\sigma_3)_{ij}
\left|p;N;ij\rangle\right.\;.
\label{D1tildestates}
\end{eqnarray}
For $n_1$ $\widetilde{D1}_\eta$-branes we then have four sectors, labeled
by the $2n_1\times 2n_1$ matrices $\Lambda =\{\mathbb{I},\Sigma_1,\Sigma_2,\Sigma_3\}$, 
where $\Sigma_i=\mathbb{I}\otimes\sigma_i$.

Let us now compute the relevant one-loop amplitudes.
The cylinder amplitude for $n_1$ 
$\widetilde{D1}_\eta$-branes is given by 
\be
{\cal A}_C  =  {n_1^2\over 2}\int_0^{\infty}{dt\over 2t}\,
  \int {d^2 p\over (2\pi)^2}
  \,\mbox{Tr}_{NS}\left(P^+_{GSO} + P^-_{GSO} + P^-_{GSO} + P^+_{GSO}\right)
  e^{-2\pi tL_0} \;,
\ee
where we have explicitly included the sum over the four sectors
with the appropriate GSO-projections;
the $\mathbb{I}$ and $\Sigma_3$ sectors contain only GSO-even states, 
and the $\Sigma_1$ and $\Sigma_2$ sectors contain only GSO-odd states.
This reduces to twice the amplitude for unprojected open NS strings
\be
{\cal A}_C =  2\cdot {n_1^2\over 2}\int_0^{\infty}{dt\over 2t}\,
  \int {d^2 p\over (2\pi)^2}\,
  e^{-\pi t(p_0^2+p_1^2)}
  = {n_1^2\over (2\pi)^3} \int_0^\infty ds
   = n_1^2 \int_0^\infty ds\, \langle T\rangle_{\widetilde{D1}_\eta}^2\;,
\ee
which reflects the fact that there is no R-R exchange in the closed
string channel, and that the ``tension'' of the $\widetilde{D1}_\eta$
is twice that of the $D1_\eta$.\footnote{This is different than the tension
of a ``wrong'' dimension brane, which is always $\sqrt{2}$ times
the tension of the same dimension brane in the other theory.}
The M\"obius amplitude is
\be 
{\cal A}_M = {n_1\over 2}\int_0^{\infty}{dt\over 2t}\,\int {d^2 p\over (2\pi)^2}
  \,\mbox{Tr}_{NS}\left(P^+_{GSO} \pm P^-_{GSO} \pm P^-_{GSO} - P^+_{GSO}\right)
  \Omega e^{-2\pi tL_0}\;,
\label{hat_omega_mobius_amplitude}
\ee
where we have included the two possible actions of $\widehat{\Omega}=\Omega\cdot(-1)^{F_L}$ 
on the different CP factors (so $\Omega$ here acts only on the Virasoro states).
The signs are explained as follows.
Recall that the operator $(-1)^{F_L}$ exchanges the D1-brane and anti-D1-brane, so
its action is equivalent to conjugation by $\Sigma_1$.
The action of $\Omega$ on the CP factors has two possibilities:
\be
\Omega: \Lambda \mapsto 
\left\{
\begin{array}{l}
\Lambda^T  \\
\Sigma_3\Lambda^T\Sigma_3 \;.
\end{array}
\right.
\ee
This follows from the requirement that $\widehat{\Omega}^2 = 1$,
and the fact that the D1-brane and anti-D1-brane are each invariant
under $\Omega$.
Combining this with the action of $(-1)^{F_L}$ we get
\be
\widehat{\Omega}: \Lambda \mapsto 
\left\{
\begin{array}{l}
\Sigma_1 \Lambda^T \Sigma_1 \\
\Sigma_2 \Lambda^T \Sigma_2 \;,
\end{array}
\right.
\label{hat_omega}
\ee
and hence the corresponding signs in (\ref{hat_omega_mobius_amplitude}).
The amplitude then reduces to
\begin{eqnarray}
{\cal A}_M &=& \pm 2\cdot{n_1\over 2}\int_0^{\infty}{dt\over 2t}\,
 \int {d^2 p\over (2\pi)^2}\,
 \mbox{Tr}_{NS}P^-_{GSO}\Omega e^{-2\pi tL_0}\nn\\
 &=& \pm n_1 \int_0^{\infty}{dt\over 2t}\,
 \int {d^2 p\over (2\pi)^2}\, e^{-\pi t(p_0^2 + p_1^2)}\nn\\
  &=& \pm {4n_1\over (2\pi)^3} \int_0^\infty ds 
   = 2n_1 \int_0^\infty ds\, 
   \langle T\rangle_{\widetilde{D1}_\eta}\langle T\rangle_{\widehat{O1}}\;,
\end{eqnarray}
where in the third equality we used the modular transformation 
$s=\pi/4t$. We conclude that
\be
\langle T\rangle_{\widehat{O1}}=\pm 2\langle T\rangle_{\widetilde{D1}_\eta}\;,
\label{D1O1tadrel}
\ee
and denote the two models by $\widehat{\Omega}_\pm$.
The tachyon tadpole can only be cancelled for $\widehat{\Omega}_-$.
The Klein bottle amplitude can be computed 
in a similar way, but we leave it out as it provides no new information.

The open string spectrum can be read off from (\ref{hat_omega}),
and the known action of $\Omega$ on the ground states.
Let us first consider only one type of $\widetilde{D1}_\eta$-brane,
say $\widetilde{D1}_+$.
Of the two GSO-even sectors only one survives, so the symmetry is $U(n_1)$
for both $\widehat{\Omega}_\pm$.
In the tadpole-free case we therefore have $U(2)$.
The GSO-odd sectors are both even under $\widehat{\Omega}_+$, 
and both odd under $\widehat{\Omega}_-$. Since the tachyon wave-function is even
the tachyon is in the symmetric representation in the first case, and
in the antisymmetric representation in the second case.
In the tadpole-free case we therefore have a $U(2)$ singlet tachyon.

By including both types of 1-brane we also get massless fermions. 
For $n_1^+$ $\widetilde{D1}_+$s and $n_1^-$ $\widetilde{D1}_-$s
the ``gauge group'' is $U(n_1^+)\times U(n_1^-)$, the tachyons
are in the symmetric and antisymmetric representations for
$\widehat{\Omega}_+$ and $\widehat{\Omega}_-$ case, respectively, 
and the fermions are in (two copies of)
the bi-fundamental representation (and its conjugate).
In particular, we have another tadpole-free theory with $n_1^+=n_1^-=1$,
that is with a CP group $U(1)\times U(1)$, and two charged massless fermions
plus their complex conjugates.

\subsection{0A$/\Omega$ and 0A$/\widehat{\Omega}$}

As in ten dimensions, both these models produce a tachyon tadpole,
but no R-R tadpole. The action of $\Omega$ interchanges the 
$(\R-,\R+)$ and $(\R+,\R-)$ sectors, so that $C_1^+$ is even
and $C_1^-$ is odd. Precisely the opposite holds for the 
action of $\widehat{\Omega}$, since it includes also $(-1)^{F_L}$,
namely $C_1^+$ is odd and $C_1^-$ is even. 
The massless NS-NS tachyon is even in both cases.

Let us begin with the Klein bottle amplitude. 
Note that the R-R sector does not contribute to this amplitude 
in the loop channel, for either $\Omega$ or $\widehat{\Omega}$, in Type 0A; 
the contributions of $C^+$ and $C^-$ cancel.
We therefore find
\begin{eqnarray}
{\cal A}_K&=&\half\int_0^{\infty}{dt\over 2t}
\int{d^2p\over(2\pi)^2}e^{-{\pi t\over 2}(p_0^2+p_1^2)}
\Tr_{NSNS}\Biggl({1+(-1)^{f+\tilde{f}}\over 2}
{\Omega} 
\Biggr)\nn\\
&=&{2\over (2\pi)^3}\int_0^{\infty}ds\,
 = \int_0^\infty ds\langle T\rangle^2_{O1}\;,
\end{eqnarray}
and exactly the same for $\widehat{\Omega}$.
We see that there is a tachyon tadpole in both models (and no R-R
tadpole, due to the diagonal GSO-projection).

Now consider adding $\widetilde{D1}_\eta$-branes.
The open strings between like-sign branes are NS strings.
There are two CP sectors, corresponding to the $\mathbb{I}$ and
$\sigma_1$ sectors of the $D1_\eta\overline{D1}_\eta$ pair in Type 0B 
(\ref{D1tildestates}).
This follows from Sen's construction of the $\widetilde{Dp}$-brane
in Type IIA(B) as a projection of a $Dp\overline{Dp}$ pair in Type IIB(A)
by $(-1)^{F_L}$ \cite{Sen:1999mg}. The projection removes the $\sigma_2$ and
$\sigma_3$ sectors.
The cylinder amplitude is therefore given by the amplitude for unprojected
NS strings
\be
 {\cal A}_C = {n_1^2\over 2}\int_0^{\infty}{dt\over 2t}
\,\int {d^2 p\over (2\pi)^2}e^{-\pi t(p_0^2+p_1^2)}
={n_1^2\over 2(2\pi)^3 }\int_0^{\infty}ds
 = n_1^2 \int_0^\infty ds\,\langle T\rangle_{\widetilde{D1}_\eta}^2\;.
\ee

To compute the M\"obius amplitude we first need to determine how $\Omega$
and $\widehat{\Omega}$ act on the open string sectors.
The action on the CP factors is standard
\be 
 \Omega\,,\widehat{\Omega}: \Lambda \mapsto \Gamma\Lambda^T\Gamma^{-1}\;,
 \quad \mbox{where} \; \Gamma=\mathbb{I} \;\;\mbox{or}\; \Sigma_2 \;.
\ee
The action on the open string states is then determined by the action 
on the ground states of the different sectors. 
In the GSO-even $\mathbb{I}$ sector the action is standard
\be
 \Omega,\widehat{\Omega}\ket{p;0;\mathbb{I}} = -i \ket{p;0;\mathbb{I}} \;,
\ee 
so the would-be vector, which is the first excited state (and the lowest
to survive the GSO projection) is odd under $\Omega$ and $\widehat{\Omega}$.
The CP group (which is now not really a gauge group, since there is no
gauge field) in both models is therefore $SO(n_1)$ for $\Gamma=\mathbb{I}$
and $Sp(n_1/2)$ for $\Gamma=\Sigma_2$.
The action on the GSO-odd sector
ground state, {\em i.e.} the massless tachyon, can be determined
using Sen's argument \cite{Sen:1999mg}:
since the $\widetilde{D1}_\eta$-brane world-sheet theory contains a term
\be
\int C_1\wedge dt \;,
\ee
the transformation property of the open string tachyon $t$ 
must be the same as that of $C_1$.
But which $C_1$ is this?
As we will review in the next section, only one type of D0-brane is consistent in 
two-dimensional
Type 0A string theory, $D0_-$ in our conventions.
Since the above term identifies a tachyonic kink as a source for the R-R
field $C_1$, {\em i.e.} a D0-brane, the R-R field must be $C_1^-$.
It follows that the tachyon is odd under $\Omega$ and even under $\widehat{\Omega}$,
\begin{eqnarray}
 \Omega \ket{p;0;\Sigma_1} &=& - \ket{p;0;\Sigma_1} \\
 \widehat{\Omega} \ket{p;0;\Sigma_1} &=&  \ket{p;0;\Sigma_1}\ .
\end{eqnarray}
Therefore the tachyon belongs
to the (anti)symmetric representation of Sp(SO) in the $\Omega$ model,
and to the (anti)symmetric representation of SO(Sp) in the $\widehat{\Omega}$ model.
The M\"obius amplitudes are given by
\begin{eqnarray}
{\cal A}_M &=& \pm {n\over 2} \int_0^{\infty}{dt\over 2t}
  \,\int {d^2 p\over (2\pi)^2}e^{-\pi t(p_0^2+p_1^2)}
  =\pm {2n_1\over (2\pi)^3 }\int_0^{\infty}ds \\
{\cal A}_{\widehat{M}} &=& \mbox{} - {\cal A}_M 
  =\mp {2n_1\over (2\pi)^3 }\int_0^{\infty}ds \;,
\label{0aomega11mb}
\end{eqnarray}
and therefore 
\begin{eqnarray}
 \langle T\rangle_{O1}  &=&  \pm 2 \langle T\rangle_{\widetilde{D1}_\eta} 
\label{O1tadpole_D1tadpole_0B}\\
 \langle T\rangle_{\widehat{O1}}  &=&  \mp 2 \langle T\rangle_{\widetilde{D1}_\eta} \;,
\label{hatO1tadpole_D1tadpole_0B}
\end{eqnarray}
where in both models the upper sign corresponds to the CP group $Sp(n_1/2)$,
and the lower sign to $SO(n_1)$. 
The two tadpole free models are therefore 0A$/\Omega$ with CP group $SO(2)$
and a single tachyon, and 0A$/\widehat{\Omega}$ with CP group 
$Sp(1)=SU(2)$ and a tachyon in the ${\bf 1}$.

In the more general situation where we add both types of
$\widetilde{D1}_\eta$-branes we get a product CP group,
{\em e.g.} $SO(n_1^+)\times SO(n_1^-)$, and also
massless fermions in bi-fundamental representations.
In particular, we can get one more tadpole free model with
$n_1^+=n_1^-=1$, no CP group,
no massless tachyons, only a single massless fermion.




\section{Dual Matrix Models}

In this section we will propose a dual matrix model for each of
the unoriented two-dimensional string theories constructed
in the previous sections, in terms of the corresponding unstable D0-brane
quantum mechanics. 
To obtain the matrix models for the unoriented theories we need to
determine the D0-brane gauge group and open string matter content
in each case. 
This is done using now standard techniques.
For each of the four models we will have two choices for
the gauge group and/or matter representation, just as in the D1-brane case.
What will be slightly less trivial is determining 
which choice for the D0-branes corresponds to which choice for the D1-branes.
Our approach will be to compare the relation between the D1 disk tadpole
and the $RP^2$ tadpole, which we will obtain by computing the D0-D1
cylinder amplitude and the D0-D0 M\"obius amplitude, with the corresponding
relation obtained in section 3 from the D1-D1 cylinder and M\"obius
amplitudes. This will fix the D0-brane open string data 
relative to the D1-brane open string data.
The results are summarized in table 3.

\subsection{Bosonic String}

We begin again with the bosonic string. 
The D0-brane corresponds to a Liouville Dirichlet boundary state
tensored with a $c=1$ (and ghost) Neumann boundary state.
The former is labeled by a pair of integers $(n,m)$,
in one-to-one correspondence with the degenerate conformal families $[V_{n,m}]$.
The open strings between a $(1,1)$ D0-brane and an $(n,m)$ D0-brane
belong to this family \cite{Zamolodchikov:2001ah}.
In particular, the physical open string spectrum on a $(1,1)$ D0-brane
contains only a tachyon (which is now truly tachyonic, since the mass is 
not shifted).
This is the D0-brane used in constructing the matrix model for the oriented bosonic
string \cite{Klebanov:2003km}. The role of the other $(n,m)$ D0-branes
is not yet clear.\footnote{For a recent proposal however 
see \cite{Ponsot:2003ss,Hosomichi:2001xc} and \cite{Alexandrov:2003un}.}
In what follows we will refer to the $(1,1)$ D0-brane simply as the D0-brane.

It has also been shown that the open strings between the D0-brane and a 
D1-brane with (continuous) parameter $\nu$ correspond to the conformal 
family $[V_{\alpha}]$, 
with $\alpha=Q/2+i\nu/2$ \cite{Zamolodchikov:2001ah}.
The conformal weights of $V_{\alpha}$ and $V_{n,m}$ are given by 
$\Delta_{\alpha}=Q^2/4-(Q-2\alpha)^2/4$ and 
$\Delta_{n,m}=Q^2/4-(nb+m/b)^2/4$, respectively. 
In the $(n,m)$ degenerate module, the null state appears at level
$nm$, so its conformal weight is 
$\Delta_{null}=Q^2/4-(nb-m/b)^2/4$. 

The cylinder amplitude for the D0-D1 strings is therefore given by
(including also the $c=1$ matter and ghost contributions):
\be
{\cal A}_{C,0-1} =  n_0n_1\int_0^{\infty}{dt\over 2t}
\int{dp_0\over 2\pi}e^{-\pi tp_0^2}e^{-{\pi t\over 2}\nu^2} \nn\\
 =  {n_0n_1\over 4\pi^{3/2}}\int_0^{\infty}{ds\over s^{1/2}}
         e^{-{\pi^2\over 2s}\nu^2}\;,
\label{D0D1cylinder}
\ee
where $n_0$ and $n_1$ are the numbers of D0-branes and D1-branes,
and we have performed the usual loop to tree channel
transformation for the cylinder $s=\pi/t$.

Turning our attention to the D0-D0 strings, let us first consider
the more general case of $(n,m)$ D0-branes.
To compute the cylinder or M\"obius amplitude for $(n,m)$ D0-branes
one needs to evaluate the trace in the conformal family $[V_{n,m}]$,
and subtract the trace in the null module beginning with the null state.
For the cylinder this gives
\be
 {\cal A}_{C,0-0} =  {n_0^2\over 2}\int_0^{\infty}{dt\over 2t}
\int{dp_0\over 2\pi}e^{-\pi tp_0^2}
\left(e^{{\pi t\over 2}(nb+m/b)^2}-e^{{\pi t\over 2}(nb-m/b)^2}\right) \;,
\ee
and for the M\"obius strip this gives \cite{Hikida:2002bt}
\be
{\cal A}_{M,0-0} = \pm n_0\half\int_0^{\infty}{dt\over 2t}
\int{dp_0\over 2\pi}e^{-\pi tp_0^2}
\left(e^{{\pi t\over 2}(nb+m/b)^2}-(-1)^{nm}
e^{{\pi t\over 2}(nb-m/b)^2}\right)\;.
\ee
The relative factor $(-1)^{nm}$ comes from the fact that the null module
begins at level $nm$, and from the action of $\Omega$ on the Virasoro
generators $\Omega L_n \Omega^{-1} = (-1)^n L_n$.
Specializing to the case $b=1$ (2d string), and $(n,m)=(1,1)$, and performing
the modular transformation $s=\pi/4t$, the M\"obius amplitude becomes
\be
{\cal A}_{M,0-0}=\pm{2n_0\over 4\pi^{3/2}}\int_0^{\infty}{ds\over s^{1/2}}
   e^{{\pi^2\over 4s}} \;.
\ee
Comparing the large $s$ behavior of this amplitude with the 
large $s$ behavior of the D0-D1 cylinder amplitude (\ref{D0D1cylinder})
we see that the tadpoles are related as 
$\langle T\rangle_{O1}=\pm 2\langle T\rangle_{D1}$,
in agreement with what we found in section 2 using the free field
computations of the D1-brane annulus and M\"obius amplitudes.

The upper sign in the M\"obius amplitude (or the $RP^2$ tadpole)
corresponds to the D0-brane gauge group $SO(n_0)$, and the lower sign to
$Sp(n_0/2)$, paralleling the choice of the D1-brane CP group. 
The open string tachyon on the D0-brane is even under $\Omega$,
and therefore transforms in the (anti)symmetric representation in
the SO(Sp) case.
In addition, there are $n_1$ scalars in the vector representation of
the D0-brane gauge group from the D0-D1 strings. Their mass is given by
\be
 m^2_{01} = {\nu^2\over 4\alpha'} \;.
\ee
These are tachyonic as well if $\nu$ is imaginary.
In the tadpole-free case we therefore get an $Sp(n_0/2)$ matrix model
with a symmetric matrix and two vectors.

\begin{table}
\begin{center}
\begin{tabular}[h]{|c|c|c|}
\hline
theory & D0-brane gauge group & D0-brane matter\\
\hline
 &&\\[-10pt]
bos/$\Omega_\pm$ & 
$
\begin{array}{l}
SO(n_0) \\[5pt]
Sp({n_0\over 2})
\end{array}
$ &
$
\begin{array}{l}
\symm + n_1\,\funda\\[5pt]
\asymm +n_1\,\funda
\end{array}
$
\\[20pt]
\hline
&&\\[-10pt]
0B/$\Omega_\pm$ &
$
\begin{array}{l}
SO(n_0) \\[5pt]
Sp({n_0\over 2})
\end{array}
$
 &
$
\begin{array}{l}
\asymm_{\,b} + 2n_1^-\,\funda_{\,b} + 2n_1^+\,\funda_{\,f} \\[5pt]
\symm_{\,b} + 2n_1^-\,\funda_{\,b} + 2n_1^+\,\funda_{\,f}
\end{array}
$\\[20pt]
\hline
&&\\[-10pt]
0B/$\widehat{\Omega}_\pm$ &
$
\begin{array}{l}
SO(n_0)\\[5pt]
Sp({n_0\over 2})
\end{array}
$
 &
$
\begin{array}{l}
\symm_{\,b} + 2n_1^-\,\funda_{\,b} + 2n_1^+\,\funda_{\,f}\\[5pt]
\asymm_{\,b} + 2n_1^-\,\funda_{\,b} + 2n_1^+\,\funda_{\,f}
\end{array}
$
\\[20pt]
\hline
&&\\[-10pt]
0A/$\Omega_\pm$ &
$
\begin{array}{l}
 U(n_0) \\[5pt]
 U(n_0) 
\end{array}
$
 &
$
\begin{array}{l}
\symm_{\,b} + n_1^-[\funda + \overline{\funda}]_b
   + n_1^+[\funda + \overline{\funda}]_f \\[5pt]
\asymm_{\,b} +  n_1^-[\funda + \overline{\funda}]_b
   + n_1^+[\funda + \overline{\funda}]_f
\end{array}
$
\\[20pt]
\hline
&&\\[-10pt]
0A/$\widehat{\Omega}_\pm$ &
$
\begin{array}{l}
 SO(n_0) \times SO(\bar{n}_0)\\[5pt]
 Sp({n_0\over 2})\times Sp({\bar{n}_0\over 2})
\end{array}
$
 &
$
\begin{array}{l}
(\funda,\funda)_b + n_1^-[(\funda,{\bf 1}) + ({\bf 1},\funda)]_b
    + n_1^+[(\funda,{\bf 1}) + ({\bf 1},\funda)]_f \\[5pt]
 (\funda,\funda)_b + n_1^-\,[(\funda,{\bf 1}) + ({\bf 1},\funda)]_b
    + n_1^+[(\funda,{\bf 1}) + ({\bf 1},\funda)]_f 
\end{array}
$
\\[20pt]
\hline
\end{tabular}
\caption{D0-brane matrix models for unoriented 2d string theories.}
\end{center}
\label{matrix_model_table}
\end{table}


\subsection{Type 0}

The Dirichlet boundary state in super-Liouville 
theory is again labeled by a pair of integers $(n,m)$.
In this case the open strings between a $(1,1)$ state and
an $(n,m)$ state can only be NS if $n-m$ is even, and R if $n-m$ is odd.
The open strings between a $(1,1)$ and an $(n,m)$ Dirichlet state,
and between a $(1,1)$ Dirichlet state and a $\nu$ Neumann state
again correspond to the conformal families of $V_{n,m}$ and $V_\alpha$,
respectively, where $\alpha=Q/2 +i\nu/2$. 
However the conformal weights of $V_{n,m}$ and $V_\alpha$ are
now given by $\Delta_{n,m}=Q^2/8-(nb+m/b)^2/8$ and 
$\Delta_{\alpha}=Q^2/8-(Q-2\alpha)^2/8$ in the NS sector,
and by $\Delta_{n,m}=Q^2/8+1/16-(nb+m/b)^2/8$ and 
$\Delta_{\alpha}=Q^2/8+1/16-(Q-2\alpha)^2/8$ in the R sector.
The null state appears at level $nm/2$ in the $(n,m)$
degenerate module in both the NS and R sectors.
As in the bosonic case, we will only be interested in the $(1,1)$ Dirichlet state.

In super-Liouville theory it turns out that the Dirichlet boundary state
is consistent only for one value of the spin structure $\eta$, namely
for $\eta = -\sgn(\mu_0)$.\footnote{Our convention for $\eta$ 
differs from that of \cite{Douglas:2003up} by a sign. Compare for example their 
equation (5.9) with our equation (\ref{cylinder_conditions}).}
This implies that there is only one type of D0-brane
in two-dimensional Type 0A or Type 0B string theory. 
For definiteness let us fix
$\mu_0>0$. Then in Type 0A there is only a $D0_-$, and in 
Type 0B there is only a $\widetilde{D0}_-$.
This is also consistent with the fact that in
Type 0A the exponential tachyon background given by the Liouville interaction
allows only the R-R field strength $F^- = dC^-_1$ to have a non-vanishing
time-independent value \cite{Douglas:2003up}.
It is also consistent with the fact that $(-1)^f$ is not a symmetry.
This operator changes the sign of $\eta$, and therefore interchanges the
two types of D-brane.
Had it been a symmetry, both types would have to exist 
(as in the ten-dimensional theories).
Of course the absence of this symmetry does not forbid the presence of
the other type of D-brane, as we see in the D1-brane case.

The matrix models for the oriented theories correspond
to the quantum mechanics of $n_0$ $\widetilde{D0}_-$-branes in Type 0B,
and $n_0$ $D0_-$-branes plus $\bar{n}_0$ $\overline{D0}_-$-branes in Type 0A
\cite{Douglas:2003up}. 
The former is therefore the Hermitian $U(n_0)$ matrix
model, and the latter is the quiver $U(n_0)\times U(\bar{n}_0)$ matrix model.
Let us now determine the corresponding matrix models for the unoriented theories.

\subsubsection{The 0B models}

The $\widetilde{D0}_-$-$\widetilde{D0}_-$
open strings are NS strings, so $n-m$ is restricted to
be even. Like the $\widetilde{D1}_\eta$-$\widetilde{D1}_\eta$ strings
in Type 0A, there are two CP sectors: the GSO-even $\mathbb{I}$ sector,
and the GSO-odd $\Sigma_1$ sector. 
The action of $\Omega$ and $\widehat{\Omega}$ on the GSO-even sector is the same 
as in the case of the Type 0A $\widetilde{D1}_\eta$-brane, so the gauge group
is $SO(n_0)$ for $\Gamma =\mathbb{I}$, and $Sp(n_0/2)$ for $\Gamma=\Sigma_2$.
To determine the action on the GSO-odd sector we will use Sen's argument again.
The $\widetilde{D0}_-$-brane world-line theory contains the term
\begin{equation}
\int C_0\wedge dt \;,
\end{equation}
which survives in the unoriented theory. Since $C_0$ is odd under $\Omega$ and even
under $\widehat{\Omega}$, we conclude that the same holds for the tachyon, namely
\begin{eqnarray}
 \Omega \ket{(n,m);0;\Sigma_1} &=& - \ket{(n,m);0;\Sigma_1} \nn\\
 \widehat{\Omega} \ket{(n,m);0;\Sigma_1} &=& \ket{(n,m);0;\Sigma_1}\;.
\label{omega_0B}
\end{eqnarray}
Therefore the tachyon is in the (anti)symmetric representation of the
Sp(SO) gauge group in the $\Omega$ model, 
and the (anti)symmetric representation of the SO(Sp) gauge group in
 the $\widehat{\Omega}$ model. 
In addition there are scalar and fermion fields in the vector representation
coming from the 0-1 strings.
What is not known at this stage is which 0-brane gauge group goes with which
1-brane CP group. To determine this, we will derive the relation between
the disk and $RP^2$ tadpoles from the 0-1 cylinder and 0-0 M\"obius amplitudes,
and compare with the relation obtained in section 3.

Let us start with the 0-1 cylinder amplitude. As there are two kinds of D1-brane,
$D1_\pm$ in the $\Omega$ model and $\widetilde{D1}_\pm$ in the $\widehat{\Omega}$
model, there are two amplitudes to consider.
The $\widetilde{D0}_-$-$D1_-$ (or $\widetilde{D1}_-$) strings are unprojected 
NS strings,
and the $\widetilde{D0}_-$-$D1_+$ (or $\widetilde{D1}_+$) strings are 
unprojected R strings.
%
The NS and R amplitudes are actually the same in two-dimensions, and given
by\footnote{We have included an extra factor of $\half$ so that the 
results we will find below are consistent with
  those in section 3. We will do so in the type 0A case as well. However,
  we do not have a clear understanding of how this factor appears from
  the open string viewpoint. In any case, what we are after is the relative sign
of the tadpoles.} 
\begin{eqnarray}
{\cal A}_{C,\widetilde{0}-1} &=& {1\over 2}n_0n_1\int_0^{\infty}{dt\over 2t}
\int{dp_0\over 2\pi}e^{-\pi tp_0^2}e^{-{\pi t\over 4}\nu^2}
={n_0n_1\over 8\pi^{3/2}}\int_0^{\infty}{ds\over s^{1/2}}
e^{-{\pi^2\over 4s}\nu^2}\label{cylinder0B}\\
{\cal A}_{C,\widetilde{0}-\widetilde{1}} &=& 2{\cal A}_{C,0-1}\;.
\label{cylinder0Btilde}
\end{eqnarray}
The factor of 2 for the $\widetilde{D0}_-$-$\widetilde{D1}_\eta$
amplitude is due to the fact 
that $\widetilde{D1}_\eta$ is a $D1_\eta\overline{D1}_\eta$ combination.

Now consider the $\widetilde{D0}_-$- $\widetilde{D0}_-$ M\"obius
amplitude. In the Liouville part, we need to sum over the states in the Verma module of
$V_{n,m}$, and subtract the sum over the states in the null submodule.
The Liouville M\"obius partition function is given by
%
\begin{eqnarray}
Z^{L}_{n,m}(t)&=& \mbox{Tr}'_N\Biggl[
\left.\langle (n,m),N;\mathbb{I}\right|P^+_{GSO}\,\Omega\,
e^{-2\pi t\left(L^{L}_0-{\hat{c}_L\over 16}\right)}
\left|(n,m),N;\mathbb{I}\rangle\right.\nn\\
&&\pm\left.\langle (n,m),N;\Sigma_1\right|P^-_{GSO}\,\Omega\,
e^{-2\pi t\left(L^{L}_0-{\hat{c}_L\over 16}\right)}
\left|(n,m),N;\Sigma_1\rangle\right.\Biggr]\;,
\label{liouvillenm0bpf}
\end{eqnarray}
where the upper and lower signs correspond to the $\Omega$ and
$\widehat{\Omega}$ cases, respectively, as follows from (\ref{omega_0B}).
%
%
The Liouville Hamiltonian
is given by (with $\hat{c}_L=1+2Q^2$) 
\begin{equation}
L^{(L)}_0-{\hat{c}_L\over 16}
=-{1\over 8}(nb+m/b)^2+N_B+N_F-{1\over 16} \;,
\end{equation}
and the primed trace denotes the subtraction of the null
submodule. 
There are two points to make about the subtraction.
First, the null submodule gets a factor of $i^{nm}$ relative to the
full Verma module. This comes from the action of $\Omega$ on the 
superconformal modes, $\Omega G_r \Omega = (-1)^r G_r$, in the NS sector.
Second, the GSO projection is reversed relative to
the full Verma module if $nm$ is odd.
These points are elaborated upon in Appendix B.  
For $(n,m)=(1,1)$ we then get
\begin{equation}
Z^{L}_{1,1}(t)=
-i\left(e^{{\pi t\over 4}(b+1/b)^2}
\pm e^{{\pi t\over 4}(b-1/b)^2}\right)
Z_{GSO+}
\mp
\left(e^{{\pi t\over 4}(b+1/b)^2}
\mp e^{{\pi t\over 4}(b-1/b)^2}\right)
Z_{GSO-}\ ,
\end{equation}
where the upper signs hold for $\Omega$ and the lower signs for $\widehat{\Omega}$,
and where we have defined
\begin{eqnarray}
Z_{GSO\pm}\equiv
\frac{\vartheta_{00}(0,it+1/2)^{1/2}
\mp\vartheta_{01}(0,it+1/2)^{1/2}}
{2e^{-i\pi/16}\eta(it+1/2)^{3/2}}\ .\label{ZGSOpm}
\end{eqnarray}

Including $\hat{c}=1$ matter and ghosts (which cancel the oscillator contributions), 
and CP factors, and taking $b\rightarrow 1$,
the M\"obius amplitudes become
\begin{eqnarray}
{\cal A}_{M,\widetilde{0}-\widetilde{0}} &=&  \pm 
{n_0\over 4\pi^{3/2}}\int_0^{\infty}{ds\over s^{1/2}}
\Biggl[-i\left(e^{{\pi^2\over 4s}}
 + 1 \right)\half(1-1)
 - \left(e^{{\pi^2\over 4s}} - 1 \right)\half(1+1)\Biggr] 
\label{0Bomega00}\\
{\cal A}_{\widehat{M},\widetilde{0}-\widetilde{0}} &=&  \pm 
{n_0\over 4\pi^{3/2}}\int_0^{\infty}{ds\over s^{1/2}}
\Biggl[-i\left(e^{{\pi^2\over 4s}}
 - 1 \right)\half(1-1)
 + \left(e^{{\pi^2\over 4s}} + 1 \right)\half(1+1)\Biggr],
\label{0Bomegahat00}
\end{eqnarray}
for the $\Omega$ and $\widehat{\Omega}$ cases, respectively.
The first of the two terms in the bracket comes from GSO-even states only,
and vanishes since there are no propagating GSO-even states. We keep it with
the factor $(1-1)$  to remind ourselves that the $1$ comes from 
tracing over NS states with a $1\cdot\Omega$, and the $-1$ comes from tracing
over NS states with a $(-1)^f \cdot\Omega$. The relative minus sign is due
to the fact that the ground state of the NS sector is odd under $(-1)^f$.

Comparing now (\ref{cylinder0Btilde}) with (\ref{0Bomegahat00}) in the 
$s\to\infty$ limit for the $\widehat{\Omega}$ case, we read off 
\begin{equation}
\langle T\rangle_{\widehat{O1}}
=\pm 2\langle T\rangle_{\widetilde{D1}_\eta}\;,
\end{equation}
in agreement with what was found in (\ref{D1O1tadrel}).
It follows that the D0-brane gauge group is $SO(n_0)$ for $\widehat{\Omega}_+$
and $Sp(n_0/2)$ for $\widehat{\Omega}_-$.


>From the large $s$ limit of the $\Omega$ amplitude we see that
\begin{equation}
\langle T\rangle_{O1}=0\ .
\end{equation}
This is consistent with the result in the previous section.
However, in order to fix the D0-brane gauge group 
we need to compare (\ref{0Bomega00})
with (\ref{0Bomegamobius}) more carefully. 
In particular, the would-be tadpole corresponds to the leading term
\be
 \pm {n_0\over 4\pi^{3/2}}\int_0^{\infty}{ds\over s^{1/2}}
 \Biggl[-i(1-1)\Biggl] \;.
\ee
As stated above, the $1$ comes from an NS trace with $1\cdot\Omega$,
and the $-1$ comes from an NS trace with $(-1)^f\cdot\Omega$.
By channel duality, these correspond to the leading contributions
of the tree amplitudes $\bra{B,-}\Delta\ket{C,+}$ and  $\bra{B,-}\Delta\ket{C,-}$,
respectively. The same holds for the $1$ and $-1$ terms in (\ref{0Bomegamobius}).
Given that the normalization of the D1-brane is
fixed by the cylinder amplitude (\ref{cylinder0B}), we conclude 
that the gauge group assignment is again $SO(n_0)$ for $\Omega_+$,
and $Sp(n_0/2)$ for $\Omega_-$.

\subsubsection{The 0A models}

Since we have fixed $\mu_0>0$, Type 0A contains only 
a $D0_-$-brane. 
Recalling that $C_1^-$  is odd under $\Omega$ and even under $\widehat{\Omega}$,
it follows that the $D0_-$-brane is invariant under
$\widehat{\Omega}$, and is mapped to the $\overline{D0}_-$-brane
under $\Omega$. We denote the invariant combination in the second
case by $\widetilde{D0}_-$. The analysis of the gauge group and matter
representation parallels the one for the 1-branes in Type 0B (except that
the roles of $\Omega$ and $\widehat{\Omega}$ are reversed).
In the $\widehat{\Omega}$ theory the gauge group is either $SO(n_0)\times SO(\bar{n}_0)$
or $Sp(n_0/2)\times Sp(\bar{n}_0/2)$, the tachyon is in the bi-fundamental
representation, and there are additional scalars and fermions in the vector
representation. In the ${\Omega}$ theory the gauge group is $U(n_0)$
(for both ${\Omega}_\pm$), and the tachyon is either in the symmetric
or antisymmetric representation. Here too there are additional scalars and 
fermions from the 0-1 strings.
We will now compute the 0-1 cylinder and 0-0 M\"obius amplitudes in order to fix
which gauge group goes with which of $\widehat{\Omega}_\pm$, and which tachyon representation
goes with which of ${\Omega}_\pm$.

The 0-1 cylinder amplitude  is exactly the
same as in the type 0B case, namely
\begin{eqnarray}
{\cal A}_{C,0-\widetilde{1}} & = & n_0n_1\,
\half\int_0^{\infty}{dt\over 2t}
\int{dp_0\over 2\pi}e^{-\pi tp_0^2}e^{-{\pi t\over 4}\nu^2}
={n_0n_1\over 8\pi^{3/2}}\int_0^{\infty}{ds\over s^{1/2}} 
e^{-{\pi^2\over 4s}\nu^2}
\label{cylinder0Aomega}\\
{\cal A}_{C,\widetilde{0}-\widetilde{1}}& = & 2 {\cal A}_{C,0-\widetilde{1}} \;,
\label{cylinder0Aomegahat}
\end{eqnarray}
for the $\widehat{\Omega}$ and $\Omega$ model, respectively.
Note that we included an extra factor of $1/2$ as
we remarked in footnote 12.

For the M\"obius amplitude we again consider first the Liouville M\"obius 
partition function for the general $(n,m)$ Dirichlet state.
In the $\widehat{\Omega}$ model it is given by 
\begin{equation}
Z^{L,\widehat{\Omega}}_{n,m}(t) = \Tr'_N
\left.\langle (n,m),N\right|P^+_{GSO}\,\Omega\,
e^{-2\pi t\left(L^{(L)}_0-{\hat{c}_L\over 16}\right)}
\left|(n,m),N\rangle\right.\;,
\end{equation}
where the primed trace denotes the subtraction of the null submodule.
In the $\Omega$ model the $\widetilde{D0}_-$-$\widetilde{D0}_-$
strings have four CP sectors, just as in (\ref{D1tildestatesfirst}).
The action of GSO and $\Omega$ on the CP factors is exactly 
the same as the action of GSO and $\widehat{\Omega}$ for the 
$\widetilde{D1}_\pm$-brane in Type 0B.
Therefore the Liouville partition function, after summing over the CP
factors, yields
\begin{equation}
Z^{L,\Omega}_{n,m}(t)=\pm 2\Tr'_N
\left.\langle (n,m),N\right|P^-_{GSO}\,\Omega\,
e^{-2\pi t\left(L^{(L)}_0-{\hat{c}_L\over 16}\right)}
\left|(n,m),N\rangle\right.\;,
\end{equation}
where the choice of sign corresponds to $\Omega_\pm$,
defined in analogy with (\ref{hat_omega}). 
The computation proceeds in parallel to
the Type 0B case. For $(n,m)=(1,1)$ one finds
\begin{eqnarray}
Z^{L,\widehat{\Omega}}_{1,1}(t)&=&
-i e^{{\pi t\over 4}(b+1/b)^2}
Z_{GSO+}+ e^{{\pi t\over 4}(b-1/b)^2}
Z_{GSO-}\ ,\\
Z^{L,{\Omega}}_{1,1}(t)&=&\pm 2
\Biggl(e^{{\pi t\over 4}(b+1/b)^2}
Z_{GSO-}+i e^{{\pi t\over 4}(b-1/b)^2}
Z_{GSO+}\Biggr)\ .
\end{eqnarray}
Including the $\hat{c}=1$ matter and ghosts, and taking $b\rightarrow 1$,
the M\"obius amplitudes become
\begin{eqnarray}
{\cal A}_{\widehat{M},0-0}&=&
\pm {n_0\over 4\pi^{3/2}}\int_0^{\infty}{ds\over s^{1/2}}\label{0aomegamb}\\
{\cal A}_{{M},\widetilde{0}-\widetilde{0}}&=&
\pm {2n_0\over 4\pi^{3/2}}\int_0^{\infty}{ds\over s^{1/2}}
e^{\pi^2\over 4s}\ ,\label{oaomegahatmb}
\end{eqnarray}
for the $\widehat{\Omega}$ and ${\Omega}$ models, respectively.
In the fomer case the sign determines the gauge group as $SO(n_0)$ or 
$Sp(n_0/2)$, whereas in the latter case the gauge group is $U(n_0)$,
and the sign determines the tachyon representation.
Comparing (\ref{0aomegamb}) with (\ref{cylinder0Aomega}), and
(\ref{oaomegahatmb}) with (\ref{cylinder0Aomegahat}) in the
$s\to\infty$ limit, one can read off that
\begin{eqnarray}
\langle T\rangle_{\widehat{O1}}
&=&\pm 2\langle T\rangle_{\widetilde{D1}_\eta} \\
\langle T\rangle_{{O1}}
&=&\pm 2\langle T\rangle_{\widetilde{D1}_\eta}\ .
\end{eqnarray}
Now comparing these with (\ref{hatO1tadpole_D1tadpole_0B}) and 
(\ref{O1tadpole_D1tadpole_0B}), we conclude
that in 0A$/\widehat{\Omega}$ the D0-brane gauge group is $SO(n_0)\times SO(\bar{n}_0)$ for
$\widehat{\Omega}_+$ and $Sp(n_0/2)\times Sp(\bar{n}_0/2)$ for $\widehat{\Omega}_-$, 
and that in 0A$/{\Omega}$ the tachyon is symmetric for ${\Omega}_+$
and antisymmetric for ${\Omega}_-$.
This is shown, together with the fundamental scalars and fermions,
in table 3.


\section{Discussion}

We have classified the possible unoriented string theories in two dimensions,
and constructed their (conjectured) dual matrix models. The tadpole-free theories are 
listed in table 1. 
Unlike RR tadpoles, 
the massless tachyon tadpole is not inconsistent,
but rather is a source of quantum corrections to the
classical (tachyon) background \cite{Fischler:ci}. 
Thus the tadpole-non-free theories are as sensible as the
tadpole-free ones, unless the quantum corrections lead to
a runaway tachyon potential.
Whereas in critical string theory it is not easy to implement the
Fischler-Susskind mechanism beyond a small number of loops,
we expect that in two dimensions
the exact quantum corrected background should be built-in
to the dual matrix models. 
It would be interesting to systematically extract the quantum shifts
of the classical background from a matrix model computation.

In the (bosonic and Type 0B) oriented case without D1-branes the dual 
matrix models are equivalent to free fermions in an inverted harmonic 
oscillator potential. The background is represented by the Fermi surface.
It is therefore conceivable that matrix models for tadpole-free theories
in general can be described by non-interacting fermions.
It is also possible that matrix models for theories with non-vanishing tadpoles
may have a description in terms of interacting fermions 
\cite{Brezin:dk,Minahan:1992bz,Kazakov:1991pt,Avan:1995sp}. 
It would be very interesting to study the unoriented matrix models 
in more detail.

\section*{Acknowledgments}
We would like to thank J.~Gomis and Y.~Nakayama for useful discussions,
and J.~Park for pointing out an error in an earlier version of the paper.
This work is supported in part by the
Israel Science Foundation under grant no.~101/01-1.


\appendix
\renewcommand{\theequation}{\Alph{section}.\arabic{equation}}
\renewcommand{\thetable}{\Alph{section}.\arabic{table}}
\setcounter{table}{0}
\section{Channel duality}

The cylinder, M\"obius strip, and Klein bottle amplitudes
can be expressed either as one-loop (open or closed) vacuum amplitudes,
or as tree level amplitudes corresponding to the propagation
of closed strings between boundary states and/or crosscap states.
We refer to the relation between the loop-channel and tree-channel 
descriptions as {\em channel duality}. In this appendix we shall
review the channel duality map for the world-sheet fermions. 
The results are summarized in tables (\ref{cylinder_duality}),
(\ref{Mobius_duality}) and (\ref{Klein_duality}),
where we translate the different spin structures of the wordlsheet
fermions into the tree and loop channels for the cylinder, M\"obius strip
and Klein bottle.
The translation tables are equally valid
for critical and non-critical superstrings,
as they only depend on the topology of the given world-sheet.

\subsection{Cylinder}

Let us begin with the cylinder.
We have to impose two boundary conditions and a periodicity
condition on the world-sheet fermions:
\be
\begin{array}{lrcl}
 \mbox{Boundary I} & \tilde{\psi}(0,\sigma^2) & = & \eta_1 \psi(0,\sigma^2) \\
 \mbox{Boundary II} & \tilde{\psi}(s,\sigma^2) & = & \eta_2 \psi(s,\sigma^2) \\
 \mbox{Periodicity} & \psi(\sigma^1,2\pi) &  = & \eta_3 \psi(\sigma^1,0) \\
          & \tilde{\psi}(\sigma^1,2\pi) & = & \eta_4 \tilde{\psi}(\sigma^1,0),
\end{array}
\label{cylinder_conditions}
\ee
where $\eta_i=\pm 1$. The four phases are not independent however.
Combining one of the boundary conditions with the periodicity condition 
applied to that boundary we infer that $\eta_3=\eta_4 \equiv \eta'$.
In the tree channel $\sigma^1$ plays the role of time and $\sigma^2$
plays the role of space. From this point of view the above conditions 
define boundary states
with spin structures $\ket{B_I,\eta_1}_{\eta'}$ and 
$\ket{B_{II},\eta_2}_{\eta'}$, where $\eta' = \pm 1$ refers to the 
R-R and NS-NS sector, respectively.
Actually, since we can multiply the left-movers relative to the 
right-movers by a phase, the amplitude only depends on the product 
$\eta_1\eta_2 \equiv \eta$, so there are only four independent spin 
structures parameterized by $\eta$ and $\eta'$.
In the loop channel the roles of $\sigma^1$ and $\sigma^2$ are interchanged,
so we now find that $\eta=\pm 1$ corresponds to the open string R
and NS sector, respectively, and $\eta' = \pm 1$ corresponds to
the insertion of $(-1)^f$ and $1$ in the open string trace, respectively.
\begin{table}[h]
\begin{center}
\begin{tabular}{|c|l|l|}
\hline
 $(\eta',\eta)$ & tree channel & loop channel \\
\hline
&&\\[-12pt]
$(+,+)$ & $\bra{B,\pm}\Delta\ket{B,\pm}_{RR}$ & NS $(-1)^f$ \\[5pt]
$(+,-)$ & $\bra{B,\pm}\Delta\ket{B,\mp}_{RR}$ & R $(-1)^f$ \\[5pt]
$(-,+)$ & $\bra{B,\pm}\Delta\ket{B,\pm}_{NSNS}$ & NS $1$ \\[5pt]
$(-,-)$ & $\bra{B,\pm}\Delta\ket{B,\mp}_{NSNS}$ & R $1$ \\[5pt]
\hline
\end{tabular}
\caption{Cylinder channel duality.}
\end{center}
\label{cylinder_duality}
\end{table}

\subsection{M\"obius strip}

In the case of the M\"obius strip we have to impose one boundary 
condition, one crosscap condition and a periodicity condition:
\be
\begin{array}{llcl}
 \mbox{Boundary} & \tilde{\psi}(0,\sigma^2) & = & \eta_1 \psi(0,\sigma^2) \\
 \mbox{Crosscap} & \psi(s,\sigma^2) & = &  
     \eta_2 \tilde{\psi}(s,\sigma^2 + \pi) \\
  &  \tilde{\psi}(s,\sigma^2) & = &  
     \eta_3 \psi(s,\sigma^2 + \pi) \\
 \mbox{Periodicity} & \psi(\sigma^1,2\pi) &  = & \eta_4 \psi(\sigma^1,0) \\
          & \tilde{\psi}(\sigma^1,2\pi) & = & \eta_5 \tilde{\psi}(\sigma^1,0)\;.
\end{array}
\label{Mobius_conditions}
\ee
As in the case of the cylinder, there are only four independent spin structures,
which one can take to be parameterized by $\eta=\eta_1\eta_2$ and 
$\eta' = \eta_4$; multiplication of the left-movers by an overall phase
fixes, say, $\eta_2=+1$,
the boundary condition then forces $\eta_5=\eta'$, and the crosscap condition
(applied twice) requires $\eta_3 = \eta'$.
In the tree channel we then have a boundary state $\ket{B,\eta_1}_\eta'$
and a crosscap state $\ket{C,\eta_2}_\eta'$, where, as before,
$\eta' = \pm 1$ corresponds to the R-R and NS-NS sector, respectively.
In the loop channel the M\"obius strip corresponds to a twisted identification
of two opposite boundaries of a rectangle. To get this we 
first double the range of $\sigma^1$ to $[0,2s]$ by copying the 
$(\sigma^1,\sigma^2)$ plane and reflecting the copy about the $\sigma^1$ axis,
and then halve the range of $\sigma^2$ to $[0,\pi]$.
In the new domain we define
\begin{eqnarray}
\Psi(\sigma^1,\sigma^2) & \equiv &
\left\{ 
\begin{array}{ll}
 \psi(\sigma^1,\sigma^2) & \quad \sigma^1 <s \\
 \tilde{\psi}(2s-\sigma^1,\sigma^2+\pi) & \quad \sigma^1>s
\end{array}
\right. \nonumber \\
\tilde{\Psi}(\sigma^1,\sigma^2) & \equiv &
\left\{ 
\begin{array}{ll}
 \tilde{\psi}(\sigma^1,\sigma^2) & \quad \sigma^1 <s \\
 \eta'\psi(2s-\sigma^1,\sigma^2+\pi) & \quad \sigma^1>s\;.
\end{array}
\right.
\label{new_fermions}
\end{eqnarray}
The conditions (\ref{Mobius_conditions}) then imply the boundary condition
\begin{eqnarray}
 \Psi(0,\sigma^2) & = & \eta \tilde{\Psi}(0,\sigma^2) \nonumber\\
 \Psi(2s,\sigma^2) & = & \eta\eta' \tilde{\Psi}(2s,\sigma^2) \nonumber
\end{eqnarray}
and the twisted periodicity condition
\begin{eqnarray}
 \Psi(\sigma^1,0) & = & \eta' \tilde{\Psi}(2s-\sigma^1,\pi) \nonumber\\
 \tilde{\Psi}(\sigma^1,0) & = & \Psi(2s-\sigma^1,\pi)\;. \nonumber
\end{eqnarray}
By redefining the left-mover by a phase $\eta$ we see that
$\eta'=\pm 1$ corresponds to the R and NS open string
sector, respectively, 
and that $\eta = \pm 1$ corresponds to the two possible insertions
$(-1)^f\cdot\Omega$ and $1\cdot\Omega$. It isn't immediately
clear which insertion corresponds to which sign of $\eta$,
but an explicit calculation of the amplitude shows that $\eta=+1$
corresponds to $(-1)^f\cdot\Omega$, and $\eta=-1$ 
to $1\cdot\Omega$ (see for example \cite{Bergman:1997gf}, where this was 
done for the critical superstring).
\begin{table}[h]
\begin{center}
\begin{tabular}{|c|l|l|}
\hline
 $(\eta',\eta)$ & tree channel & loop channel \\
\hline
&&\\[-12pt]
$(+,+)$ & $\bra{B,\pm}\Delta\ket{C,\pm}_{RR}$ & R $(-1)^f\cdot\Omega$ \\[5pt]
$(+,-)$ & $\bra{B,\pm}\Delta\ket{C,\mp}_{RR}$ & R $1\cdot\Omega$ \\[5pt]
$(-,+)$ & $\bra{B,\pm}\Delta\ket{C,\pm}_{NSNS}$ & NS $(-1)^f\cdot\Omega$ \\[5pt]
$(-,-)$ & $\bra{B,\pm}\Delta\ket{C,\mp}_{NSNS}$ & NS $1 \cdot\Omega$ \\[5pt]
\hline
\end{tabular}
\caption{M\"obius strip channel duality.}
\end{center}
\label{Mobius_duality}
\end{table}

\subsection{Klein bottle}

The Klein bottle requires two crosscap conditions
and a periodicity condition:
\be
\begin{array}{llcl}
 \mbox{Crosscap I} & \psi(0,\sigma^2) & = & \eta_1\tilde{\psi}(0,\sigma^2 + \pi) \\
   &   \tilde{\psi}(0,\sigma^2) &=& \eta_2 \psi(0,\sigma^2+\pi) \\
 \mbox{Crosscap II} & \psi(s,\sigma^2) & = &  
     \eta_3 \tilde{\psi}(s,\sigma^2 + \pi) \\
  &  \tilde{\psi}(s,\sigma^2) & = &  
     \eta_4 \psi(s,\sigma^2 + \pi) \\
 \mbox{Periodicity} & \psi(\sigma^1,2\pi) &  = & \eta_5 \psi(\sigma^1,0) \\
          & \tilde{\psi}(\sigma^1,2\pi) & = & \eta_6 \tilde{\psi}(\sigma^1,0)\;.
\end{array}
\label{KB_conditions}
\ee
Once again, there are only four independent spin structures, which we take
to be parameterized by $\eta=\eta_1\eta_3$ and $\eta'=\eta_5$.
We can use the freedom to redefine the left-mover by an overall phase
to fix $\eta_3=+1$. Then the constraints from applying the two crosscap
conditions twice give $\eta_2=\eta\eta'$ and $\eta_4=\eta_6=\eta'$.
The above conditions then define two crosscap states 
$\ket{C_I,\eta_1}_\eta'$ and $\ket{C_{II},\eta_3}_\eta'$ in the tree channel.
For the loop channel we again have to work in the domain
$\sigma^1\in[0,2s]$ and $\sigma^2\in[0,\pi]$.
The fermions in the new domain are then defined precisely as in 
(\ref{new_fermions}). We now get a periodicity condition in the $\sigma^1$
direction,
\begin{eqnarray}
\Psi(0,\sigma^2) & = & \eta\Psi(2s,\sigma^2) \nonumber\\
 \tilde{\Psi}(0,\sigma^2) & = & \eta\tilde{\Psi}(2s,\sigma^2)\;,
\end{eqnarray}
and a twisted periodicity condition in the $\sigma^2$ direction,
\begin{eqnarray}
\Psi(\sigma^1,0) & = & \eta'\tilde{\Psi}(2s-\sigma^1,\pi) \nonumber\\
 \tilde{\Psi}(\sigma^1,0) & = & \Psi(2s-\sigma^1,\pi)\;.
\end{eqnarray}
Therefore $\eta=\pm 1$ corresponds to the (closed string) R-R and NS-NS sector,
respectively, and $\eta'=\pm 1$ corresponds to the insertion of
$((-1)^f + (-1)^{\tilde{f}})\cdot\Omega$ and 
$(1+(-1)^{f+\tilde{f}})\cdot\Omega$ in the closed string trace,
respectively.\footnote{The precise correspondence is again 
verified by an explicit calculation \cite{Bergman:1997gf}.
It may seem odd that the 
Klein bottle has only four spin structures rather
than eight (like the torus),
corresponding in the loop channel to the choice of R-R or NS-NS sector,
and to the separate insertions $\Omega$, $(-1)^f\Omega$, $(-1)^{\tilde{f}}\Omega$
or $(-1)^{f+\tilde{f}}\Omega$. This is because $\Omega$ relates the left-moving
state to the right-moving state, and in particular identifies the left and 
right-moving spin structures.}
\begin{table}[h]
\begin{center}
\begin{tabular}{|c|l|l|}
\hline
 $(\eta',\eta)$ & tree channel & loop channel \\
\hline
&&\\[-12pt]
$(+,+)$ & $\bra{C,\pm}\Delta\ket{C,\pm}_{RR}$ 
        & R-R $((-1)^f+(-1)^{\tilde{f}})\cdot\Omega$ \\[5pt]
$(+,-)$ & $\bra{C,\pm}\Delta\ket{C,\mp}_{RR}$ 
        & NS-NS $((-1)^f+(-1)^{\tilde{f}})\cdot\Omega$ \\[5pt]
$(-,+)$ & $\bra{C,\pm}\Delta\ket{C,\pm}_{NSNS}$ 
        & R-R $(1+(-1)^{f+\tilde{f}})\cdot\Omega$ \\[5pt]
$(-,-)$ & $\bra{C,\pm}\Delta\ket{C,\mp}_{NSNS}$ 
        & NS-NS $(1+(-1)^{f+\tilde{f}})\cdot\Omega$ \\[5pt]
\hline
\end{tabular}
\caption{Klein bottle channel duality.}
\end{center}
\label{Klein_duality}
\end{table}


\section{Some relevant calculations}

In this appendix, we show some details of the computations. 
  
\subsection{The GSO projected character of degenerate representation
  on $\mathbb{RP}^2$}

In section 4, we computed the character of degenerate representation
on $\mathbb{RP}^2$. 
Here we will elaborate the details of the calculation in the type 0B
case. The generalization to the type 0A case is straightforward.    
Let us define the GSO projected character on $\mathbb{RP}^2$ by
\begin{eqnarray}
\chi_{\mathbb{I}}^{GSO\pm}(\Delta)
&\equiv&
\Tr_{\mathbb{I}}\left(P^{\pm}_{GSO}\,\Omega\,
q^{\Delta+N_B+N_F}\right)\ ,\nn\\ 
\chi_{\Sigma_1}^{GSO\pm}(\Delta) 
&\equiv&
\Tr_{\Sigma_1}\left(P^{\pm}_{GSO}\,\Omega\,q^{\Delta+N_B+N_F}
\right)\ .
\end{eqnarray}
The trace is over the states in the $\mathbb{I}$ sector for 
$\chi_{\mathbb{I}}$,
and those in the $\Sigma_1$ sector for 
$\chi_{\Sigma_1}$, where two CP sectors are defined in section 4. 
Recall the difference of the $\Omega$-action on the ground state (thus
the highest weight state) in each sector.  
In order to understand how to subtract the null
submodule, let us expand the characters and examine the low order
terms (we consider the $(1,1)$ degenerate module, but the
generalization to the $(n,m)$ case is straightforward):
\begin{eqnarray}
\chi_{\mathbb{I}}^{GSO\pm}(\Delta)
&=&-\half i\Biggl[(1-iq^{1/2}+\cdots)\pm(-1-iq^{1/2}+\cdots)\Biggr]
q^{\Delta}\ ,\nn\\
\chi_{\Sigma_1}^{GSO\pm}(\Delta)
&=&\half\Biggl[(1-iq^{1/2}+\cdots)\pm(-1-iq^{1/2}+\cdots)\Biggr]
q^{\Delta}\ .
\end{eqnarray}
The terms of $q^{1/2}$ corresponds to the null states in the case 
$(n,m)=(1,1)$. 
We would like to compare these with the character of the null module
\begin{eqnarray}
\chi_{\mathbb{I}}^{GSO\pm}(\Delta+1/2)
&=&-\half iq^{1/2}\Biggl[(1-iq^{1/2}+\cdots)\pm(-1-iq^{1/2}+\cdots)
\Biggr]q^{\Delta}\ ,\nn\\
\chi_{\Sigma_1}^{GSO\pm}(\Delta+1/2)
&=&\half q^{1/2}\Biggl[(1-iq^{1/2}+\cdots)\pm(-1-iq^{1/2}+\cdots)
\Biggr]q^{\Delta}\ .
\end{eqnarray}
Then we see that the subtraction should be taken as
\begin{equation}
\chi_{\mathbb{I},\Sigma_1}^{GSO\pm}(\Delta)
+i\chi_{\mathbb{I},\Sigma_1}^{GSO\mp}(\Delta+1/2)\ .
\end{equation}
Note the following two points; We need a factor of $i$, and  
the GSO projection must be reversed in the subtraction. For the 
$(n,m)$ case, the factor of $i$ be replaced by $i^{nm}$, and the GSO 
projection is reversed for $nm$ odd and the same for $nm$ even. 

Then the Liouville partition function as defined in 
(\ref{liouvillenm0bpf}) yields 
\begin{eqnarray}
Z^{L}_{n,m}(t)&=&
-i\left\{e^{{\pi t\over 4}(nb+m/b)2}Z_{GSO,+}
+i^{nm}e^{{\pi t\over 4}(nb-m/b)2}Z_{GSO,-}^{n,m}\right\}
\nn\\
&&\hspace{1.5cm}
\mp\left\{e^{{\pi t\over 4}(nb+m/b)2}Z_{GSO,-}
+i^{nm}e^{{\pi t\over 4}(nb-m/b)2}Z_{GSO,+}^{n,m}\right\}\ .
\label{Lnm0Bpart}
\end{eqnarray}
where $Z_{GSO,\pm}$ are defined in eq.(\ref{ZGSOpm}) and 
we have defined 
\begin{equation}
Z_{GSO,\pm}^{n,m}\equiv\frac{\vartheta_{00}(0,it+1/2)^{1/2} 
\mp(-1)^{nm}\vartheta_{01}(0,it+1/2)^{1/2}}
{2e^{-i\pi/16}\eta(it+1/2)^{3/2}}\ .
\end{equation}
%
\subsection{Boundary and cross-cap states}

The boundary states in the super Liouville theory were constructed in   
\cite{Fukuda:2002bv, Ahn:2002ev}, by solving the boundary bootstrap
equations obtained by making use of the degenerate conformal field and
imposing the Cardy condition, generalizing the work of
\cite{Fateev:2000ik,  Zamolodchikov:2001ah}. 
In this appendix, we will not perform the computation of one-point
functions. Instead we use the result obtained in 
\cite{Fukuda:2002bv, Ahn:2002ev} as an input for constructing, in
particular, the cross-cap states.\footnote{The crosscap state
for the two-dimensional bosonic string was constructed in \cite{Hikida:2002bt}.}
We only consider the NS-NS sector, which is relevant for our
analysis. 

The building block of the boundary and cross-cap states is the
Ishibashi states that have the following properties,
\begin{eqnarray}
\mbox{}_{NSNS}\langle B;P',\pm|\Delta
|B;P,\pm\rangle_{NSNS}&=&\delta(P'+P)
{\vartheta_{00}(0,is/\pi)^{1/2}\over\eta(is/\pi)^{3/2}}\ ,\\
\mbox{}_{NSNS}\langle B;P',\pm|\Delta
|B;P,\mp\rangle_{NSNS}&=&\delta(P'+P)
{\vartheta_{01}(0,is/\pi)^{1/2}\over\eta(is/\pi)^{3/2}}\ ,\\
\mbox{}_{NSNS}\langle B;P',\pm|\Delta
|C;P,\pm\rangle_{NSNS}&=&\delta(P'+P)
{\vartheta_{00}(0,is/\pi+1/2)^{1/2}\over
e^{-i\pi/16}\eta(is/\pi+1/2)^{3/2}}\ ,\\
\mbox{}_{NSNS}\langle B;P',\pm|\Delta
|C;P,\mp\rangle_{NSNS}&=&\delta(P'+P)
{\vartheta_{01}(0,is/\pi+1/2)^{1/2}\over
e^{-i\pi/16}\eta(is/\pi+1/2)^{3/2}}\ ,\\
\mbox{}_{NSNS}\langle C;P',\pm|\Delta
|C;P,\pm\rangle_{NSNS}&=&\delta(P'+P)
{\vartheta_{00}(0,is/\pi)^{1/2}\over\eta(is/\pi)^{3/2}}\ ,\\
\mbox{}_{NSNS}\langle C;P',\pm|\Delta
|C;P,\mp\rangle_{NSNS}&=&\delta(P'+P)
{\vartheta_{01}(0,is/\pi)^{1/2}\over\eta(is/\pi)^{3/2}}\ ,
\end{eqnarray}
where we have defined $\Delta\equiv e^{-s(N+\widetilde{N}-1/8)}$.
$N=N_B+N_F$ is the sum of bosonic and fermionic parts of the
Virasoro level operator and likewise for $\widetilde{N}$.

The NS-NS boundary state for the $(1,1)$ Dirichlet brane is 
\begin{equation}
|B_{1,1},-\rangle_{NSNS}=\int_0^{\infty}dP\,\Psi_{1,1}^{NS}(P)
|B;P,-\rangle_{NSNS}\ ,
\end{equation}
where the boundary state wave-function $\Psi_{1,1}^{NS}(P)$ is given by 
\cite{Fukuda:2002bv, Ahn:2002ev}
\begin{equation}
\Psi_{1,1}^{NS}(P)=4(\pi\mu\gamma(bQ/2))^{-iP/b}
\frac{\Gamma(1+iPb)\Gamma(1+iP/b)}{-2i\pi P}
\sinh(\pi Pb)\sinh(\pi P/b)\ .
\end{equation}
The Liouville partition function for the 0-1 open string is given by
\begin{eqnarray}
Z^{annulus}_{\alpha=Q/2+i\nu/2}
&\equiv&\mu_1\Tr_{NS}\,e^{-2\pi t\left(L_0^{(L)}-{\hat{c}\over 16}
\right)}
=\mu_1e^{-{\pi t\over 4}\nu2}
{\vartheta_{00}(0,it)^{1/2}\over \eta(it)^{3/2}}\nn\\
&=&\mu_1\left({\pi\over s}\right)^{1/2}e^{-{\pi2\over 4s}\nu2}
{\vartheta_{00}(0,is/\pi)^{1/2}\over \eta(is/\pi)^{3/2}}\nn\\
&=&2\mu_1\int_0^{\infty}dP e^{-sP2}
{\vartheta_{00}(0,is/\pi)^{1/2}\over \eta(is/\pi)^{3/2}}
\cos(\pi\nu P)\ ,
\label{01annulus}
\end{eqnarray}
where $\mu_1=2$ for $\widetilde{D0}-\widetilde{D1}$ in both type 0A and 0B, and 
$\mu_1=1$ for $D0-\widetilde{D1}$ in type 0A, and $\widetilde{D0}-D1$  
in type 0B. This determines the NS-NS boundary state for the Neumann
brane to be
\begin{equation}
|B_{\nu},-\rangle_{NSNS}=\int_0^{\infty}dP\,\Psi_{\nu}^{NS}(P)
|B;P,-\rangle_{NSNS}\ ,
\end{equation}
where the boundary state wave-function $\Psi_{\nu}^{NS}(P)$ is given by 
\cite{Fukuda:2002bv, Ahn:2002ev}
\begin{equation}
\Psi_{\nu}^{NS}(P)=\mu_1(\pi\mu\gamma(bQ/2))^{-iP/b}
\frac{\Gamma(1+iPb)\Gamma(1+iP/b)}{-i \pi  P}
\cos(\pi\nu P)\ .
\end{equation}
%


The Liouville partition function (\ref{Lnm0Bpart}) on $\mathbb{RP}^2$
for type 0B takes, in the tree channel, the form
\begin{equation}
Z^{L}_{1,1}=\left({\pi\over 2s}\right)^{1/2}\Bigg[
-i\left(e^{{\pi2\over 16s}(b+1/b)2}
\pm e^{{\pi2\over 16s}(b-1/b)2}\right)
Z^{BC-}
\mp\left(e^{{\pi2\over 16s}(b+1/b)2}
\mp e^{{\pi2\over 16s}(b-1/b)2}\right)
Z^{BC+}
\Biggr]\ ,
\end{equation}
where $s=\pi/4t$ and we have defined
\begin{equation}
Z^{BC\pm}\equiv\frac{\vartheta_{01}(0,is/\pi+1/2)^{1/2}
\pm e^{-\pi i/4}\vartheta_{00}(0,is/\pi+1/2)^{1/2}}
{2e^{-3i\pi/16}\eta(is/\pi+1/2)^{3/2}}\ .
\end{equation}
%
%
%
It can be further rewritten as
\begin{eqnarray}
Z^{L}_{1,1}&=&\sqrt{2}\int_0^{\infty}dPe^{-sP2}
\Bigg[
-i\left(\cosh(\pi(b+1/b)P/2)
\pm \cosh(\pi(b-1/b)P/2)\right)
Z^{BC-}\nn\\
&&\hspace{1cm}\mp\left(\cosh(\pi(b+1/b)P/2)
\mp \cosh(\pi(b-1/b)P/2)\right)
Z^{BC+}
\Biggr]\ .
\label{Z11mobius}
\end{eqnarray}
Then the NS-NS cross-cap state can be read off as 
(up to irrelevant phase factors which will be cancelled upon the
inclusion of the matter and ghost parts)
\begin{eqnarray}
|C\rangle_{NSNS}&=&\half\int_0^{\infty}dP\Biggl[\Psi^{NS}_{C-}(P)
\Biggl(|C;P,+\rangle_{NSNS}-|C;P,-\rangle_{NSNS}\Biggr)
\nn\\
&&\hspace{2cm}+\Psi^{NS}_{C+}(P)
\Biggl(|C;P,+\rangle_{NSNS}+|C;P,-\rangle_{NSNS}\Biggr)\Biggr]\ .
\end{eqnarray}
The cross-cap state wave-function $\Psi^{NS}_{C\pm}(P)$ is found
to be  
\begin{eqnarray}
\Psi^{NS}_{C-}(P)&=&-2\sqrt{2}i(\pi\mu\gamma(bQ/2))^{-iP/b}
\frac{\Gamma(1+iPb)\Gamma(1+iP/b)}{-2i\pi P}
\sinh(\pi Pb/2)\sinh(\pi P/2b)\ ,\nn\\
\Psi^{NS}_{C+}(P)&=&2\sqrt{2}(\pi\mu\gamma(bQ/2))^{-iP/b}
\frac{\Gamma(1+iPb)\Gamma(1+iP/b)}{-2i\pi P}
\cosh(\pi Pb/2)\cosh(\pi P/2b)\ ,\nn
\end{eqnarray}
for the upper sign (0B/$\hat{\Omega}$ model), and 
\begin{eqnarray}
\Psi^{NS}_{C-}(P)&=&-2\sqrt{2}i(\pi\mu\gamma(bQ/2))^{-iP/b}
\frac{\Gamma(1+iPb)\Gamma(1+iP/b)}{-2i\pi P}
\cosh(\pi Pb/2)\cosh(\pi P/2b)\ ,\nn\\
\Psi^{NS}_{C+}(P)&=&-2\sqrt{2}(\pi\mu\gamma(bQ/2))^{-iP/b}
\frac{\Gamma(1+iPb)\Gamma(1+iP/b)}{-2i\pi P)}
\sinh(\pi Pb/2)\sinh(\pi P/2b)\ ,\nn
\end{eqnarray}
for the lower sign (0B/$\Omega$ model).

\end{document}